\documentclass[10pt,conference]{IEEEtran}
\IEEEoverridecommandlockouts
\usepackage{cite}
\usepackage[T1]{fontenc}
\usepackage{graphicx}
\usepackage{amssymb}
\usepackage{amsmath}
\usepackage{amsthm}
\usepackage[caption=false,font=footnotesize]{subfig}
\usepackage{stfloats}
\usepackage{float}
\usepackage{microtype}
\usepackage{balance}
\usepackage{xcolor}
\usepackage{algorithm}
\usepackage{algorithmicx}
\usepackage{algpseudocode}
\usepackage[acronym]{glossaries}
\usepackage{url}
\usepackage{balance}
\usepackage{mathtools, nccmath}
\usepackage{booktabs}
\usepackage{multirow}
\usepackage[utf8]{inputenc}
\usepackage{enumitem}
\usepackage{tikz}
\usetikzlibrary{matrix}
\usetikzlibrary{external}

\usepackage{pgfplots}
\pgfplotsset{compat=1.18}
\usepgfplotslibrary{groupplots}

\pgfplotsset{
  my/axis base/.style={
    axis background/.style={fill=white},
    axis x line*=bottom,
    axis y line*=left,
    xmajorgrids,
    ymajorgrids,
    xlabel style={font=\color{white!15!black}},
    ylabel style={font=\color{white!15!black}},
    xlabel near ticks,
    ylabel near ticks,
    legend style={
      at={(0.97,0.03)}, anchor=south east,
      legend cell align=left, align=left,
      draw=white!15!black, font=\small,
      outer sep=0pt
    },
    trim axis left, trim axis right
  },
  my/range/.style n args={4}{
    xmin=#1, xmax=#2, ymin=#3, ymax=#4
  },
}

\pgfplotsset{
  toacdf/.style      ={my/axis base, my/range={0}{2}{0}{1}},
  doacdf/.style      ={my/axis base, my/range={0}{120}{0}{1}},
  mpcdelaysep/.style ={my/axis base, my/range={0}{0.5}{0}{100}},
  mpcazsep/.style ={my/axis base, my/range={-2.5}{25}{0}{100}},
  mpcelsep/.style ={my/axis base, my/range={-5}{50}{0}{100}},
}

\pgfplotsset{
  ablation/base/.style={
    my/axis base,
    tick label style={font=\small},
    xmin=-22, xmax=33,
    xtick={-20,-10,0,10,20,30},
    xticklabels={{-20},{-10},{0},{10},{20},{$\infty$}},
    xlabel style={font=\normalsize},
    ylabel style={font=\normalsize},
    yminorticks=true,
    yminorgrids,
    legend style={legend cell align=left, align=left},
    legend pos=north east
  },
  ablationtoa/.style ={ablation/base, my/range={-22}{33}{0.0225}{2.25}},
  ablationdoa/.style ={ablation/base, my/range={-22}{33}{0.5}{50}},
}

\algrenewcommand\algorithmicindent{0.7em}

\usepackage{amssymb}
\usepackage{amsfonts}
\usepackage{mathrsfs}
\usepackage{xspace}
\usepackage{bm}
\usepackage{upgreek}

\newcommand{\safemath}[2]{\newcommand{#1}{\ensuremath{#2}\xspace}}

\safemath{\bma}{\mathbf{a}}
\safemath{\bmb}{\mathbf{b}}
\safemath{\bmc}{\mathbf{c}}
\safemath{\bmd}{\mathbf{d}}
\safemath{\bme}{\mathbf{e}}
\safemath{\bmf}{\mathbf{f}}
\safemath{\bmg}{\mathbf{g}}
\safemath{\bmh}{\mathbf{h}}
\safemath{\bmi}{\mathbf{i}}
\safemath{\bmj}{\mathbf{j}}
\safemath{\bmk}{\mathbf{k}}
\safemath{\bml}{\mathbf{l}}
\safemath{\bmm}{\mathbf{m}}
\safemath{\bmn}{\mathbf{n}}
\safemath{\bmo}{\mathbf{o}}
\safemath{\bmp}{\mathbf{p}}
\safemath{\bmq}{\mathbf{q}}
\safemath{\bmr}{\mathbf{r}}
\safemath{\bms}{\mathbf{s}}
\safemath{\bmt}{\mathbf{t}}
\safemath{\bmu}{\mathbf{u}}
\safemath{\bmv}{\mathbf{v}}
\safemath{\bmw}{\mathbf{w}}
\safemath{\bmx}{\mathbf{x}}
\safemath{\bmy}{\mathbf{y}}
\safemath{\bmz}{\mathbf{z}}
\safemath{\bmzero}{\mathbf{0}}
\safemath{\bmone}{\mathbf{1}}
\safemath{\Bell}{\ensuremath{\boldsymbol\ell}}

\bmdefine{\biad}{a}
\bmdefine{\bibd}{b}
\bmdefine{\bicd}{c}
\bmdefine{\bidd}{d}
\bmdefine{\bied}{e}
\bmdefine{\bifd}{f}
\bmdefine{\bigd}{g}
\bmdefine{\bihd}{h}
\bmdefine{\biid}{i}
\bmdefine{\bijd}{j}
\bmdefine{\bikd}{k}
\bmdefine{\bild}{l}
\bmdefine{\bimd}{m}
\bmdefine{\bind}{n}
\bmdefine{\biod}{o}
\bmdefine{\bipd}{p}
\bmdefine{\biqd}{q}
\bmdefine{\bird}{r}
\bmdefine{\bisd}{s}
\bmdefine{\bitd}{t}
\bmdefine{\biud}{u}
\bmdefine{\bivd}{v}
\bmdefine{\biwd}{w}
\bmdefine{\bixd}{x}
\bmdefine{\biyd}{y}
\bmdefine{\bizd}{z}

\bmdefine{\bixid}{\xi}
\bmdefine{\bilambdad}{\lambda}
\bmdefine{\bimud}{\mu}
\bmdefine{\bithetad}{\theta}
\bmdefine{\biphid}{\phi}
\bmdefine{\bideltad}{\delta}

\safemath{\bmia}{\biad}
\safemath{\bmib}{\bibd}
\safemath{\bmic}{\bicd}
\safemath{\bmid}{\bidd}
\safemath{\bmie}{\bied}
\safemath{\bmif}{\bifd}
\safemath{\bmig}{\bigd}
\safemath{\bmih}{\bihd}
\safemath{\bmii}{\biid}
\safemath{\bmij}{\bijd}
\safemath{\bmik}{\bikd}
\safemath{\bmil}{\bild}
\safemath{\bmim}{\bimd}
\safemath{\bmin}{\bind}
\safemath{\bmio}{\biod}
\safemath{\bmip}{\bipd}
\safemath{\bmiq}{\biqd}
\safemath{\bmir}{\bird}
\safemath{\bmis}{\bisd}
\safemath{\bmit}{\bitd}
\safemath{\bmiu}{\biud}
\safemath{\bmiv}{\bivd}
\safemath{\bmiw}{\biwd}
\safemath{\bmix}{\bixd}
\safemath{\bmiy}{\biyd}
\safemath{\bmiz}{\bizd}

\safemath{\bmxi}{\bixid}
\safemath{\bmlambda}{\bilambdad}
\safemath{\bmmu}{\bimud}
\safemath{\bmtheta}{\bithetad}
\safemath{\bmphi}{\biphid}
\safemath{\bmdelta}{\bideltad}

\safemath{\bA}{\mathbf{A}}
\safemath{\bB}{\mathbf{B}}
\safemath{\bC}{\mathbf{C}}
\safemath{\bD}{\mathbf{D}}
\safemath{\bE}{\mathbf{E}}
\safemath{\bF}{\mathbf{F}}
\safemath{\bG}{\mathbf{G}}
\safemath{\bH}{\mathbf{H}}
\safemath{\bI}{\mathbf{I}}
\safemath{\bJ}{\mathbf{J}}
\safemath{\bK}{\mathbf{K}}
\safemath{\bL}{\mathbf{L}}
\safemath{\bM}{\mathbf{M}}
\safemath{\bN}{\mathbf{N}}
\safemath{\bO}{\mathbf{O}}
\safemath{\bP}{\mathbf{P}}
\safemath{\bQ}{\mathbf{Q}}
\safemath{\bR}{\mathbf{R}}
\safemath{\bS}{\mathbf{S}}
\safemath{\bT}{\mathbf{T}}
\safemath{\bU}{\mathbf{U}}
\safemath{\bV}{\mathbf{V}}
\safemath{\bW}{\mathbf{W}}
\safemath{\bX}{\mathbf{X}}
\safemath{\bY}{\mathbf{Y}}
\safemath{\bZ}{\mathbf{Z}}

\safemath{\bZero}{\mathbf{0}}
\safemath{\bOne}{\mathbf{1}}
\safemath{\bDelta}{\mathbf{\Delta}}
\safemath{\bLambda}{\mathbf{\UpLambda}}
\safemath{\bPhi}{\mathbf{\Upphi}}
\safemath{\bSigma}{\mathbf{\Upsigma}}
\safemath{\bOmega}{\mathbf{\Upomega}}
\safemath{\bTheta}{\mathbf{\Uptheta}}

\bmdefine{\biAd}{A}
\bmdefine{\biBd}{B}
\bmdefine{\biCd}{C}
\bmdefine{\biDd}{D}
\bmdefine{\biEd}{E}
\bmdefine{\biFd}{F}
\bmdefine{\biGd}{G}
\bmdefine{\biHd}{H}
\bmdefine{\biId}{I}
\bmdefine{\biJd}{J}
\bmdefine{\biKd}{K}
\bmdefine{\biLd}{L}
\bmdefine{\biMd}{M}
\bmdefine{\biOd}{N}
\bmdefine{\biPd}{O}
\bmdefine{\biQd}{P}
\bmdefine{\biRd}{R}
\bmdefine{\biSd}{S}
\bmdefine{\biTd}{T}
\bmdefine{\biUd}{U}
\bmdefine{\biVd}{V}
\bmdefine{\biWd}{W}
\bmdefine{\biXd}{X}
\bmdefine{\biYd}{Y}
\bmdefine{\biZd}{Z}

\bmdefine{\biDelta}{\Delta}
\bmdefine{\biLambda}{\Lambda}
\bmdefine{\biPhi}{\Phi}
\bmdefine{\biSigma}{\Sigma}
\bmdefine{\biOmega}{\Omega}
\bmdefine{\biTheta}{\Theta}

\safemath{\bimA}{\biAd}
\safemath{\bimB}{\biBd}
\safemath{\bimC}{\biCd}
\safemath{\bimD}{\biDd}
\safemath{\bimE}{\biEd}
\safemath{\bimF}{\biFd}
\safemath{\bimG}{\biGd}
\safemath{\bimH}{\biHd}
\safemath{\bimI}{\biId}
\safemath{\bimJ}{\biJd}
\safemath{\bimK}{\biKd}
\safemath{\bimL}{\biLd}
\safemath{\bimM}{\biMd}
\safemath{\bimN}{\biNd}
\safemath{\bimO}{\biOd}
\safemath{\bimP}{\biPd}
\safemath{\bimQ}{\biQd}
\safemath{\bimR}{\biRd}
\safemath{\bimS}{\biSd}
\safemath{\bimT}{\biTd}
\safemath{\bimU}{\biUd}
\safemath{\bimV}{\biVd}
\safemath{\bimW}{\biWd}
\safemath{\bimX}{\biXd}
\safemath{\bimY}{\biYd}
\safemath{\bimZ}{\biZd}

\safemath{\bimDelta}{\biDelta}
\safemath{\bimLambda}{\biLambda}
\safemath{\bimPhi}{\biPhi}
\safemath{\bimSigma}{\biSigma}
\safemath{\bimOmega}{\biOmega}
\safemath{\bimTheta}{\biTheta}

\safemath{\setA}{\mathcal{A}}
\safemath{\setB}{\mathcal{B}}
\safemath{\setC}{\mathcal{C}}
\safemath{\setD}{\mathcal{D}}
\safemath{\setE}{\mathcal{E}}
\safemath{\setF}{\mathcal{F}}
\safemath{\setG}{\mathcal{G}}
\safemath{\setH}{\mathcal{H}}
\safemath{\setI}{\mathcal{I}}
\safemath{\setJ}{\mathcal{J}}
\safemath{\setK}{\mathcal{K}}
\safemath{\setL}{\mathcal{L}}
\safemath{\setM}{\mathcal{M}}
\safemath{\setN}{\mathcal{N}}
\safemath{\setO}{\mathcal{O}}
\safemath{\setP}{\mathcal{P}}
\safemath{\setQ}{\mathcal{Q}}
\safemath{\setR}{\mathcal{R}}
\safemath{\setS}{\mathcal{S}}
\safemath{\setT}{\mathcal{T}}
\safemath{\setU}{\mathcal{U}}
\safemath{\setV}{\mathcal{V}}
\safemath{\setW}{\mathcal{W}}
\safemath{\setX}{\mathcal{X}}
\safemath{\setY}{\mathcal{Y}}
\safemath{\setZ}{\mathcal{Z}}
\safemath{\emptySet}{\varnothing}

\safemath{\colA}{\mathscr{A}}
\safemath{\colB}{\mathscr{B}}
\safemath{\colC}{\mathscr{C}}
\safemath{\colD}{\mathscr{D}}
\safemath{\colE}{\mathscr{E}}
\safemath{\colF}{\mathscr{F}}
\safemath{\colG}{\mathscr{G}}
\safemath{\colH}{\mathscr{H}}
\safemath{\colI}{\mathscr{I}}
\safemath{\colJ}{\mathscr{J}}
\safemath{\colK}{\mathscr{K}}
\safemath{\colL}{\mathscr{L}}
\safemath{\colM}{\mathscr{M}}
\safemath{\colN}{\mathscr{N}}
\safemath{\colO}{\mathscr{O}}
\safemath{\colP}{\mathscr{P}}
\safemath{\colQ}{\mathscr{Q}}
\safemath{\colR}{\mathscr{R}}
\safemath{\colS}{\mathscr{S}}
\safemath{\colT}{\mathscr{T}}
\safemath{\colU}{\mathscr{U}}
\safemath{\colV}{\mathscr{V}}
\safemath{\colW}{\mathscr{W}}
\safemath{\colX}{\mathscr{X}}
\safemath{\colY}{\mathscr{Y}}
\safemath{\colZ}{\mathscr{Z}}

\safemath{\opA}{\mathbb{A}}
\safemath{\opB}{\mathbb{B}}
\safemath{\opC}{\mathbb{C}}
\safemath{\opD}{\mathbb{D}}
\safemath{\opE}{\mathbb{E}}
\safemath{\opF}{\mathbb{F}}
\safemath{\opG}{\mathbb{G}}
\safemath{\opH}{\mathbb{H}}
\safemath{\opI}{\mathbb{I}}
\safemath{\opJ}{\mathbb{J}}
\safemath{\opK}{\mathbb{K}}
\safemath{\opL}{\mathbb{L}}
\safemath{\opM}{\mathbb{M}}
\safemath{\opN}{\mathbb{N}}
\safemath{\opO}{\mathbb{O}}
\safemath{\opP}{\mathbb{P}}
\safemath{\opQ}{\mathbb{Q}}
\safemath{\opR}{\mathbb{R}}
\safemath{\opS}{\mathbb{S}}
\safemath{\opT}{\mathbb{T}}
\safemath{\opU}{\mathbb{U}}
\safemath{\opV}{\mathbb{V}}
\safemath{\opW}{\mathbb{W}}
\safemath{\opX}{\mathbb{X}}
\safemath{\opY}{\mathbb{Y}}
\safemath{\opZ}{\mathbb{Z}}
\safemath{\opZero}{\mathbb{O}}
\safemath{\identityop}{\opI}

\safemath{\veca}{\bma}
\safemath{\vecb}{\bmb}
\safemath{\vecc}{\bmc}
\safemath{\vecd}{\bmd}
\safemath{\vece}{\bme}
\safemath{\vecf}{\bmf}
\safemath{\vecg}{\bmg}
\safemath{\vech}{\bmh}
\safemath{\veci}{\bmi}
\safemath{\vecj}{\bmj}
\safemath{\veck}{\bmk}
\safemath{\vecl}{\bml}
\safemath{\vecm}{\bmm}
\safemath{\vecn}{\bmn}
\safemath{\veco}{\bmo}
\safemath{\vecp}{\bmp}
\safemath{\vecq}{\bmq}
\safemath{\vecr}{\bmr}
\safemath{\vecs}{\bms}
\safemath{\vect}{\bmt}
\safemath{\vecu}{\bmu}
\safemath{\vecv}{\bmv}
\safemath{\vecw}{\bmw}
\safemath{\vecx}{\bmx}
\safemath{\vecy}{\bmy}
\safemath{\vecz}{\bmz}

\safemath{\veczero}{\bmzero}
\safemath{\vecone}{\bmone}
\safemath{\vecxi}{\bmxi}
\safemath{\veclambda}{\bmlambda}
\safemath{\vecmu}{\bmmu}
\safemath{\vectheta}{\bmtheta}
\safemath{\vecphi}{\bmphi}
\safemath{\vecdelta}{\bmdelta}

\safemath{\matA}{\bA}
\safemath{\matB}{\bB}
\safemath{\matC}{\bC}
\safemath{\matD}{\bD}
\safemath{\matE}{\bE}
\safemath{\matF}{\bF}
\safemath{\matG}{\bG}
\safemath{\matH}{\bH}
\safemath{\matI}{\bI}
\safemath{\matJ}{\bJ}
\safemath{\matK}{\bK}
\safemath{\matL}{\bL}
\safemath{\matM}{\bM}
\safemath{\matN}{\bN}
\safemath{\matO}{\bO}
\safemath{\matP}{\bP}
\safemath{\matQ}{\bQ}
\safemath{\matR}{\bR}
\safemath{\matS}{\bS}
\safemath{\matT}{\bT}
\safemath{\matU}{\bU}
\safemath{\matV}{\bV}
\safemath{\matW}{\bW}
\safemath{\matX}{\bX}
\safemath{\matY}{\bY}
\safemath{\matZ}{\bZ}
\safemath{\matzero}{\bmzero}

\safemath{\matDelta}{\bDelta}
\safemath{\matLambda}{\bLambda}
\safemath{\matPhi}{\bPhi}
\safemath{\matSigma}{\bSigma}
\safemath{\matOmega}{\bOmega}
\safemath{\matTheta}{\bTheta}

\safemath{\matidentity}{\matI}
\safemath{\matone}{\matO}

\safemath{\rnda}{A}
\safemath{\rndb}{B}
\safemath{\rndc}{C}
\safemath{\rndd}{D}
\safemath{\rnde}{E}
\safemath{\rndf}{F}
\safemath{\rndg}{G}
\safemath{\rndh}{H}
\safemath{\rndi}{I}
\safemath{\rndj}{J}
\safemath{\rndk}{K}
\safemath{\rndl}{L}
\safemath{\rndm}{M}
\safemath{\rndn}{N}
\safemath{\rndo}{O}
\safemath{\rndp}{P}
\safemath{\rndq}{Q}
\safemath{\rndr}{R}
\safemath{\rnds}{S}
\safemath{\rndt}{T}
\safemath{\rndu}{U}
\safemath{\rndv}{V}
\safemath{\rndw}{W}
\safemath{\rndx}{X}
\safemath{\rndy}{Y}
\safemath{\rndz}{Z}

\safemath{\rveca}{\bimA}
\safemath{\rvecb}{\bimB}
\safemath{\rvecc}{\bimC}
\safemath{\rvecd}{\bimD}
\safemath{\rvece}{\bimE}
\safemath{\rvecf}{\bimF}
\safemath{\rvecg}{\bimG}
\safemath{\rvech}{\bimH}
\safemath{\rveci}{\bimI}
\safemath{\rvecj}{\bimJ}
\safemath{\rveck}{\bimK}
\safemath{\rvecl}{\bimL}
\safemath{\rvecm}{\bimM}
\safemath{\rvecn}{\bimN}
\safemath{\rveco}{\bomO}
\safemath{\rvecp}{\bimP}
\safemath{\rvecq}{\bimQ}
\safemath{\rvecr}{\bimR}
\safemath{\rvecs}{\bimS}
\safemath{\rvect}{\bimT}
\safemath{\rvecu}{\bimU}
\safemath{\rvecv}{\bimV}
\safemath{\rvecw}{\bimW}
\safemath{\rvecx}{\bimX}
\safemath{\rvecy}{\bimY}
\safemath{\rvecz}{\bimZ}

\safemath{\rvecxi}{\bmxi}
\safemath{\rveclambda}{\bmlambda}
\safemath{\rvecmu}{\bmmu}
\safemath{\rvectheta}{\bmtheta}
\safemath{\rvecphi}{\bmphi}

\safemath{\rmatA}{\bimA}
\safemath{\rmatB}{\bimB}
\safemath{\rmatC}{\bimC}
\safemath{\rmatD}{\bimD}
\safemath{\rmatE}{\bimE}
\safemath{\rmatF}{\bimF}
\safemath{\rmatG}{\bimG}
\safemath{\rmatH}{\bimH}
\safemath{\rmatI}{\bimI}
\safemath{\rmatJ}{\bimJ}
\safemath{\rmatK}{\bimK}
\safemath{\rmatL}{\bimL}
\safemath{\rmatM}{\bimM}
\safemath{\rmatN}{\bimN}
\safemath{\rmatO}{\bimO}
\safemath{\rmatP}{\bimP}
\safemath{\rmatQ}{\bimQ}
\safemath{\rmatR}{\bimR}
\safemath{\rmatS}{\bimS}
\safemath{\rmatT}{\bimT}
\safemath{\rmatU}{\bimU}
\safemath{\rmatV}{\bimV}
\safemath{\rmatW}{\bimW}
\safemath{\rmatX}{\bimX}
\safemath{\rmatY}{\bimY}
\safemath{\rmatZ}{\bimZ}

\safemath{\rmatDelta}{\bimDelta}
\safemath{\rmatLambda}{\bimLambda}
\safemath{\rmatPhi}{\bimPhi}
\safemath{\rmatSigma}{\bimSigma}
\safemath{\rmatOmega}{\bimOmega}
\safemath{\rmatTheta}{\bimTheta}

\usepackage{amssymb}
\usepackage{amsfonts}
\usepackage{mathrsfs}
\usepackage{xspace}
\usepackage{bm}
\usepackage{fancyref}
\usepackage{textcomp}

\usepackage{multirow}
\usepackage{stmaryrd}

\newenvironment{textbmatrix}{	\setlength{\arraycolsep}{2.5pt}%
								\left[\begin{matrix}}{\end{matrix}\right]%
								\raisebox{0.08ex}{\vphantom{M}}}

\def\be{\begin{equation}}
\def\ee{\end{equation}}
\def\een{\nonumber \end{equation}}
\def\mat{\begin{bmatrix}}
\def\emat{\end{bmatrix}}
\def\btm{\begin{textbmatrix}}
\def\etm{\end{textbmatrix}}

\def\ba#1\ea{\begin{align}#1\end{align}}
\def\bas#1\eas{\begin{align*}#1\end{align*}}
\def\bs#1\es{\begin{split}#1\end{split}}
\def\bg#1\eg{\begin{gather}#1\end{gather}}
\def\bml#1\eml{\begin{multline}#1\end{multline}}
\def\bi#1\ei{\begin{itemize}#1\end{itemize}}

\newcommand{\lefto}{\mathopen{}\left}

\DeclareMathOperator{\rank}{rank}			

\newcommand{\orth}{\perp}					
\newcommand{\abs}[1]{\lefto\lvert#1\right\rvert}		

\newcommand{\vecnorm}[1]{\lefto\lVert#1\right\rVert}		
\newcommand{\frobnorm}[1]{\vecnorm{#1}_{\text{F}}}	
\newcommand{\conj}[1]{\ensuremath{#1^{*}}} 	
\newcommand{\tp}[1]{\ensuremath{#1^{T}}} 		
\newcommand{\herm}[1]{\ensuremath{#1^{H}}} 	
\newcommand{\est}[1]{\ensuremath{\hat{#1}}} 	

\safemath{\dirac}{\delta}					
\safemath{\krond}{\dirac}					

\safemath{\upto}{\uparrow}
\safemath{\downto}{\downarrow}
\safemath{\iu}{j}							
\safemath{\ev}{\lambda}						
\safemath{\hilseqspace}{l^{2}}				
\newcommand{\banachfunspace}[1]{\setL^{#1}}	
\safemath{\hilfunspace}{\banachfunspace{2}}	
\newcommand{\floor}[1]{\lfloor #1 \rfloor}
\newcommand{\ceil}[1]{\lceil #1 \rceil}

\safemath{\SNR}{\textit{SNR}} 				
\safemath{\PAR}{\textit{PAR}} 				
\safemath{\No}{N_0}							
\safemath{\Es}{E_s}							
\safemath{\Eb}{E_b}							
\safemath{\EbNo}{\frac{\Eb}{\No}}
\safemath{\EsNo}{\frac{\Es}{\No}}

\DeclareMathOperator{\CHop}{\ensuremath{\opH}} 
\safemath{\tvir}{\rndh_{\CHop}}				
\safemath{\tvtf}{\rndl_{\CHop}}				
\safemath{\spf}{\rnds_{\CHop}}				
\safemath{\bff}{H_{\CHop}}					

\safemath{\ircf}{r_{h}}						
\safemath{\tftvcf}{r_{s}}					
\safemath{\tfcf}{r_{l}}						
\safemath{\bfcf}{r_{H}}						

\safemath{\tcorr}{c_h}						
\safemath{\scf}{c_{s}}						
\safemath{\tfcorr}{c_{l}}					
\safemath{\fcorr}{c_{H}}						

\safemath{\mi}{I}							
\safemath{\capacity}{C}						

\safemath{\normal}{\mathcal{N}}			
\safemath{\jpg}{\mathcal{CN}}			
\safemath{\mchain}{\leftrightarrow}		

\safemath{\dB}{\,\mathrm{dB}}
\safemath{\dBm}{\,\mathrm{dBm}}
\safemath{\Hz}{\,\mathrm{Hz}}
\safemath{\kHz}{\,\mathrm{kHz}}
\safemath{\MHz}{\,\mathrm{MHz}}
\safemath{\GHz}{\,\mathrm{GHz}}
\safemath{\s}{\,\mathrm{s}}
\safemath{\ms}{\,\mathrm{ms}}
\safemath{\mus}{\,\mathrm{\text{\textmu}s}}
\safemath{\ns}{\,\mathrm{ns}}
\safemath{\ps}{\,\mathrm{ps}}
\safemath{\meter}{\,\mathrm{m}}
\safemath{\mm}{\,\mathrm{mm}}
\safemath{\cm}{\,\mathrm{cm}}
\safemath{\m}{\,\mathrm{m}}
\safemath{\W}{\,\mathrm{W}}
\safemath{\mW}{\, \mathrm{mW}}
\safemath{\J}{\,\mathrm{J}}
\safemath{\K}{\,\mathrm{K}}
\safemath{\bit}{\,\mathrm{bit}}
\safemath{\nat}{\,\mathrm{nat}}

\safemath{\define}{\triangleq}			

\safemath{\equivalent}{\sim}
\safemath{\distas}{\sim}					
\safemath{\sdiff}{\Delta}				

\safemath{\reals}{\mathbb{R}}
\safemath{\positivereals}{\reals_{+}}
\safemath{\integers}{\mathbb{Z}}
\safemath{\posint}{\integers_{+}}
\safemath{\naturals}{\mathbb{N}}
\safemath{\posnaturals}{\naturals_{+}}
\safemath{\complexset}{\mathbb{C}}
\safemath{\rationals}{\mathbb{Q}}

\newcommand*{\fancyrefapplabelprefix}{app}		
\newcommand*{\fancyrefthmlabelprefix}{thm}		
\newcommand*{\fancyreflemlabelprefix}{lem}		
\newcommand*{\fancyrefcorlabelprefix}{cor}		
\newcommand*{\fancyrefdeflabelprefix}{def}		
\newcommand*{\fancyrefproplabelprefix}{prop}		
\newcommand*{\fancyrefexmpllabelprefix}{exmpl}
\newcommand*{\fancyrefalglabelprefix}{alg}		
\newcommand*{\fancyreftbllabelprefix}{tbl}		

\frefformat{vario}{\fancyrefseclabelprefix}{Sec.~#1}
\frefformat{vario}{\fancyrefthmlabelprefix}{Thm.~#1}
\frefformat{vario}{\fancyreftbllabelprefix}{Tbl.~#1}
\frefformat{vario}{\fancyreflemlabelprefix}{Lem.~#1}
\frefformat{vario}{\fancyrefcorlabelprefix}{Corr.~#1}
\frefformat{vario}{\fancyrefdeflabelprefix}{Def.~#1}
\frefformat{vario}{\fancyreffiglabelprefix}{Fig.~#1}
\frefformat{vario}{\fancyrefapplabelprefix}{App.~#1}
\frefformat{vario}{\fancyrefeqlabelprefix}{(#1)}
\frefformat{vario}{\fancyrefproplabelprefix}{Prop.~#1}
\frefformat{vario}{\fancyrefexmpllabelprefix}{Ex.~#1}
\frefformat{vario}{\fancyrefalglabelprefix}{Alg.~#1}

\safemath{\dictab}{[\,\dicta\,\,\dictb\,]}

\safemath{\ysig}{\bmy}
\safemath{\ysighat}{\hat{\ysig}}
\safemath{\ysigdim}{M}
\safemath{\xsig}{\bmx}
\safemath{\xsigdim}{N}
\safemath{\nx}{n_x}
\safemath{\zsig}{\bmz}
\safemath{\zsigdim}{\ysigdim}
\safemath{\rsig}{\bmr}
\safemath{\Adict}{\bA}
\safemath{\Adicttilde}{\widetilde{\Adict}}
\safemath{\Adictdim}{\outputdim\times\xsigdim}
\safemath{\avec}{\bma}
\safemath{\avectilde}{\tilde{\avec}}
\safemath{\Bdict}{\bB}
\safemath{\Bdicttilde}{\widetilde{\Bdict}}
\safemath{\Cdict}{\bC}
\safemath{\cvec}{\bmc}
\safemath{\Ddict}{\bD}
\safemath{\Ddictdim}{\ysigdim\times\xsigdim}
\safemath{\dvec}{\bmd}
\safemath{\Ddicttilde}{\widetilde{\bD}}
\safemath{\Bonb}{\bB}
\safemath{\bvec}{\bmb}
\safemath{\Bonbdim}{\ysigdim\times\ysigdim}
\safemath{\noise}{\bmn}
\safemath{\noisedim}{\ysigim}
\safemath{\err}{\bme}
\safemath{\errdim}{\ysigdim}
\safemath{\errset}{\setE}
\safemath{\nerr}{n_e}
\safemath{\delop}{\bP_\errset}
\safemath{\delopc}{\bP_{{\errset}^c}}

\safemath{\cplxi}{\imath}
\safemath{\cplxj}{\jmath}

\safemath{\dict}{\matD}
\safemath{\inputdim}{N}		
\safemath{\outputdim}{M}		
\safemath{\sparsity}{S}	
\safemath{\inputdimA}{{N_a}}	
\safemath{\inputdimB}{{N_b}}	
\safemath{\elemA}{{n_a}}	
\safemath{\elemB}{{n_b}}	
\safemath{\resA}{\matR_a}	
\safemath{\resB}{\matR_b}	
\safemath{\subD}{\matS} 
\safemath{\subA}{\matS_a} 
\safemath{\subB}{\matS_b} 
\safemath{\dicta}{\matA} 	
\safemath{\dictb}{\matB} 	
\safemath{\hollowS}{H}
\safemath{\hollowA}{H_a}
\safemath{\hollowB}{H_b}
\safemath{\cross}{Z}
\safemath{\coh}{\mu_d}			
\safemath{\coha}{\mu_a}			
\safemath{\cohb}{\mu_b}			
\safemath{\mubs}{\nu}	
\safemath{\cohm}{\mu_m} 
\safemath{\dictset}{\setD}	
\safemath{\dictsetp}{\dictset(\coh,\coha,\cohb)}	
\safemath{\dictsetgen}{\dictset_\text{gen}}
\safemath{\dictsetgenp}{\dictsetgen(\coh)}
\safemath{\dictsetonb}{\dictset_\text{onb}}
\safemath{\dictsetonbp}{\dictsetonb(\coh)}

\safemath{\leftside}{U}
\safemath{\rightsideA}{R_a}
\safemath{\rightsideB}{R_b}

\safemath{\indexS}{\setI_S} 

\safemath{\na}{n_a}			
\safemath{\nb}{n_b}			
\safemath{\coeffa}{p_i}	
\safemath{\coeffb}{q_j}	
\safemath{\seta}{\setP}		
\safemath{\setb}{\setQ}     
\safemath{\setw}{\setW}	
\safemath{\setz}{\setZ}	
\safemath{\cola}{\veca}		
\safemath{\colb}{\vecb}		
\safemath{\cold}{\vecd}		
\safemath{\inputvec}{\vecx} 	
\safemath{\error}{\vece}	
\safemath{\noiseout}{\vecz} 	
\safemath{\inputvecel}{x}
\safemath{\inputveca}{\vecx_a}
\safemath{\inputvecb}{\vecx_b}
\safemath{\outputvec}{\vecy}	
\safemath{\lambdamin}{\lambda_{\mathrm{min}}}

\safemath{\elltwo}{\ell_2}
\safemath{\ellone}{\ell_1}
\safemath{\ellzero}{\ell_0}
\safemath{\ellinf}{\ell_\infty}
\safemath{\ellinftilde}{\ell_{\widetilde\infty}}
\safemath{\licard}{Z(\coh,\coha,\cohb)}
\safemath{\xsol}{\hat{x}}
\safemath{\xbord}{x_b}		
\safemath{\xstat}{x_s}		
\safemath{\xstatLone}{\tilde{x}_s}
\safemath{\order}{\mathcal{O}} 
\safemath{\scales}{\Theta} 
\safemath{\ones}{\mathbf{1}} 
\safemath{\zeroes}{\mathbf{0}} 
\safemath{\thlone}{\kappa(\coh,\cohb)} 
\safemath{\constoneA}{\delta} 
\safemath{\constoneB}{\epsilon} 
\safemath{\nlarge}{L}				   
\safemath{\sumlarge}{S_\nlarge}
\safemath{\maxlarger}{P_\nlarge}	   
\safemath{\Pzero}{\textrm{P0}}	
\safemath{\Pone}{\textrm{P1}}
\safemath{\vecfir}{\vecw}			 
\safemath{\vecsec}{\vecz}
\safemath{\elvecfir}{w}              
\safemath{\elvecsec}{z}				 
\safemath{\nlargefir}{n}
\safemath{\normout}{\gamma}
\safemath{\auxfun}{h}
\safemath{\supp}{\textrm{supp}}

\safemath{\indexa}{\ell}
\safemath{\indexb}{r}
\safemath{\indexc}{i}
\safemath{\indexd}{j}

\safemath{\project}{P}

\safemath{\firstslotset}{\setU_1}  
\safemath{\secondslotset}{\setU_2} 
\safemath{\randomset}{\setS} 

\safemath{\Tran}{\textnormal{T}}
\safemath{\Herm}{\textnormal{H}}

\newcommand*{\fancyreflstlabelprefix}{lst}
\fancyrefaddcaptions{english}{%
  \providecommand*{\freflstname}{Listing}%
}
\frefformat{vario}{\fancyreflstlabelprefix}{%
  \freflstname\fancyrefdefaultspacing#1#3%
}

\begin{document}

\title{Interference and Multipath Resilient ToA Estimation}

\author{%
    \IEEEauthorblockN{António Barros and Christoph Studer}\\[0.0cm]
    	\IEEEauthorblockA{\em ETH Zurich, Switzerland; email: amaiabarros@ethz.ch and studer@ethz.ch}	
    \thanks{The work was supported in part by the European Commission within the context of the project 6G-REFERENCE (6G Hardware Enablers for Cell Free Coherent Communications and Sensing), funded under EU Horizon Europe under Grant 101139155.}\\[-0.5cm]
}
\maketitle

\glsresetall
\begin{abstract}
We present a computationally-efficient algorithm for time-of-arrival (ToA) estimation that is robust under multipath propagation and strong interference. Our algorithm leverages multiple receive antennas to combine adaptive spatial filtering with autodifferentiation in order to super-resolve the tap of the first-arriving path at low computational complexity and without requiring model-order estimation. 
We use simulations with ray-traced indoor propagation channels to demonstrate significant performance improvements over conventional correlation-based ToA estimation methods and subspace techniques such as JADE.
\end{abstract}

\glsresetall
\section{Introduction}

Time-of-arrival (ToA) estimation is fundamental to wireless communication systems, where it enables synchronization mechanisms such as frame-start detection or wireless time transfer \cite{Wu2011Clock}, and to positioning systems, where it is critical to determine the time-of-flight of a ranging signal \cite{Guvenc2009TOA}. 
Emerging applications and deployment scenarios are imposing stricter requirements for ToA estimation: communication networks are expected to (i) become more dense, increasing the likelihood of mutual interference \cite{Tataria20216G}, and (ii) require tighter time synchronization to enable functions such as distributed beamforming in cell-free architectures \cite{Mudumbai2009Beamforming, Holtom2024DiscoBeam}. At the same time, both communication and positioning systems are expected to operate indoors and in dense urban environments, where the impact of multipath propagation on ToA estimation is more severe.
These trends call for highly accurate ToA estimation methods that remain reliable in interference-prone, multipath-rich environments without prior knowledge of the number of interference sources or propagation paths. 
Existing estimation techniques, however, require model-order information, have a prohibitive computational complexity, degrade under strong interference, and/or suffer from multipath-induced bias. 

\subsection{Contributions}

In order to address the limitation of existing ToA estimation methods, we propose a computationally efficient generalized likelihood-ratio test (GLRT)-based algorithm that leverages multiple receive antennas. Our method (i) is resilient to interference and multipath and (ii) achieves super-resolution with no prior knowledge of the number of propagation paths. 
To arrive at a computationally-efficient implementation, we leverage automatic differentiation with JAX~\cite{Bradbury2018JAX}, which enables sub-$12$\,ms processing latency on general-purpose hardware. To validate the robustness and accuracy of our algorithm, we perform simulations using ray-traced wireless channels for an indoor factory environment at a carrier frequency of $15$\,GHz.

\subsection{Related Work}
\label{subsec:related_work}

While numerous ToA estimation techniques exist in the literature, especially for frame-start detection \cite{Schmidl1997Sync, Nasir2016TimingSync}, none jointly address interference, multipath propagation, and high-resolution estimation with tractable computational complexity.

For frame-start detection, reference \cite{Bliss2010TemporalSync} proposes a GLRT-based approach for multi-antenna systems in frequency-selective channels with interference. However, interference is modeled as Gaussian noise, which might poorly model interference, and the algorithm complexity is prohibitive. 
Building on the same approach, reference \cite{Hiltunen2015TemporalSync} derives a reduced-complexity variant for frequency flat channels, which does not extend to frequency-selective channels. 
The JASS algorithm in \cite{Marti2025JammerResilience} proposes a spatial filtering approach to tackle jamming with tractable complexity, but only for frequency-flat channels. 
In addition, none of the above methods account for subsample timing offsets, which inevitably results in insufficient accuracy for clock synchronization and positioning.

For highly accurate ToA estimation, spectral estimation methods---namely subspace-based methods and compressive sensing methods---have been proposed to decompose a channel estimate into its multipath components (MPCs) and estimate their individual delays~\cite{Kazaz2022DelayEstimation}.
Subspace-based methods include extrema-searching techniques (e.g., MUSIC \cite{Schmidt1986MUSIC}), polynomial-rooting techniques (e.g., root-MUSIC \cite{Rao1989RootMUSIC}), and matrix-shifting techniques (e.g., ESPRIT \cite{Roy1989ESPRIT} or JADE \cite{VanDerVeen1997JADE}). 
Such methods offer high resolution at manageable complexity, but require model-order knowledge and are known (i) to incur severe errors if the model-order is underestimated or (ii) produce spurious estimates if the model-order is overestimated---this complicates their deployment in real systems. 
Compressive-sensing-based methods exploit multipath delay sparsity to avoid explicit model-order selection. Prominent methods include on-grid basis pursuit denoising~\cite{Chen1998BasisPursuit,Fyhn2013CompressiveSensing}, gridless atomic-norm minimization~\cite{Candes2014SuperResolution,Li2022AtomicNorm}, and covariance-fitting~\cite{Stoica2011SPICE,Park2018SPICE}. Despite their super-resolution capabilities and automatic model-order selection, such methods typically require prohibitive complexity, which precludes them from real-time application.

\subsection{Notation}

Bold lowercase letters denote vectors (e.g., $\veca$), bold uppercase letters represent matrices (e.g., $\matA$), lowercase letters denote scalars (e.g., $a$), and uppercase letters represent algorithm, system, and model parameters (e.g., $A$). 
For a vector $\bma$, $a_n$ corresponds to the $n$th entry. 
The symbol $\odot$ denotes the Hadamard product, $\bF$ the unitary discrete Fourier transform (DFT) matrix, and $\matidentity_A$ the $A\times A$ identity. 
Conjugate, transpose, and conjugate transpose are denoted by $(\cdot)^*$, $(\cdot)^\Tran$, and $(\cdot)^\Herm$, respectively. 
The Euclidean norm is $\|\cdot\|_2$ and the Frobenius norm $\|\cdot\|_F$. The operators $\lfloor \cdot \rfloor$, $\lceil \cdot \rceil$, and $\lfloor \cdot \rceil$ denote floor, ceiling, and closest integer, respectively. For a scalar function $f\!\left(\cdot\right)$, $f'\!\left(\cdot\right)$ denotes the derivative  and $f''\!\left(\cdot\right)$ the second derivative.

\section{System Model}
\label{sec:system_model}

Our ToA estimation algorithm is designed for multi-antenna receivers that receive a known signal coming from a primary single-antenna transmitter. 
We assume that the primary transmitter transmits a $K$-length synchronization sequence $\bms_\textnormal{S} \in \opC^{K}$ and that the receiver is equipped with an $M$-antenna uniform rectangular array (URA).\footnote{A generalization to other geometries is possible but not discussed further.} The synchronization sequence is pulse-shaped with a known filter $g_\textnormal{S}(t)$, $t \in \opR$, resulting in the baseband transmit signal
\begin{align}
    x_\textnormal{S}(t) = \sum_{k=0}^{K-1} s_k g_\textnormal{S}(t - kT_\text{sym}),
    \label{eq:tx_sync_sig}
\end{align}
where $T_\text{sym}$ denotes the symbol period, and $g_\textnormal{S}(t)$ has a support limited to $t\in\left[0, RT_\textnormal{sym}\right]$, i.e., it spans $R$ symbols. Without loss of generality, we assume that a single secondary transmitter simultaneously broadcasts an unknown interference signal $x_\textnormal{I}(t)$, $t \in \opR$.
We also assume that an oversampling factor $P$ is used at the receiver. We denote $T_s = T_\text{sym} / P$ as the sampling period, such that the sampled synchronization and interference signals, delayed by $\tau$, are given, respectively, by
\begin{align}
    x_{\textnormal{S},\tau}[\ell] = x_\textnormal{S}(\ell T_s - \tau),\,\,x_{\textnormal{I},\tau}[\ell] = x_\textnormal{I}(\ell T_s - \tau), \, \ell \in \opN_0.
    \label{eq:discretized_sig} 
\end{align}

In what follows, we consider a frequency-selective block-fading channel model, in which all channel coefficients remain constant over the entire support of the synchronization signal. Thus, we model the discrete-time received baseband signal as
\begin{align}
    \bmy[\ell] = \, & \sum_{n=1}^{L_\textnormal{S}} \bma_{\textnormal{S},n}\beta_{\textnormal{S},n}x_{\textnormal{S},\tau_{\textnormal{S},n}}[\ell] \\
    & + \sum_{n=1}^{L_\textnormal{I}} \bma_{\textnormal{I},n}\beta_{\textnormal{I},n}x_{\textnormal{I},\tau_{\textnormal{I},n}}[\ell] + \bmz_\ell,
    \label{eq:discretized_signal_model}
\end{align}
where the first $L_\textnormal{S}$ terms correspond to the MPCs of the synchronization signal, the next $L_\textnormal{I}$ terms correspond to the MPCs of the interfering signal, and $\bmz_\ell \sim \jpg (\bmzero, \No\matidentity_{M})$ models i.i.d.\ complex-valued circularly-symmetric Gaussian noise with per-entry variance $\No$. 
The column vectors $\bma_{\textnormal{S},n} \in \opC^{M}$ and $\bma_{\textnormal{I},n} \in \opC^{M}$ 
correspond to the array response vectors of the $n$th MPC for the synchronization and interference signals, respectively, and $\beta_{\textnormal{S},n} \in \opC$, $\beta_{\textnormal{I},n} \in \opC$ denote their complex fading gains. 
The propagation delay of the $n$th MPC of the synchronization signal is $\tau_{\textnormal{S},n} \in \opR^+$ and the propagation delay of the $n$th MPC of the interference signal is $\tau_{\textnormal{I},n} \in \opR^+$.

The receiver has an observation window of size of $N = P(K+R-1)+N_\textnormal{P}$ samples that fits the sampled synchronization signal with extra margin for time alignment, given by $N_\textnormal{P}$. At time index $\ell$, the receiver aggregates all recorded samples for all antennas in the matrix
\begin{align}
    \bY_{\ell} = \left[\bmy[\ell], \bmy[\ell + 1], \dots, \bmy[\ell + N - 1]\right],
    \label{eq:rx_sig_mat}
\end{align}
which can be compactly represented as 
\begin{align}
    \bY_{\ell} =\,& \sum_{n=1}^{L_\textnormal{S}} \bma_{\textnormal{S},n}\beta_{\textnormal{S},n}\tp{\bmx_{\textnormal{S},\tau'_{\textnormal{S},n}}} + \sum_{n=1}^{L_\textnormal{I}} \bma_{\textnormal{I},n}\beta_{\textnormal{I},n}\tp{\bmx_{\textnormal{I},\tau'_{\textnormal{I},n}}} + \bZ,
    \label{eq:win_sig_model_time}
\end{align} 
where $\bZ = \left[\bmz_0, \bmz_1, \dots \bmz_{N-1}\right] \in \opC^{M \times N}$, $\bmx_{\textnormal{S},\tau'_{\textnormal{S},n}} \in \opC^{N}$ and $\bmx_{\textnormal{I},\tau'_{\textnormal{I},n}} \in \opC^{N}$ are column vectors given by
\begin{align}
    \bmx_{\textnormal{S},\tau'_{\textnormal{S},n}} &= \tp{\left[x_{\textnormal{S},\tau'_{\textnormal{S},n}}[0], x_{\textnormal{S},\tau'_{\textnormal{S},n}}[1], \dots, x_{\textnormal{S},\tau'_{\textnormal{S},n}}[N - 1]\right]} , 
    \label{eq:vec_sync_sig} \\
    \bmx_{\textnormal{I},\tau'_{\textnormal{I},n}} &= \tp{\left[x_{\textnormal{I},\tau'_{\textnormal{I},n}}[0], x_{\textnormal{I},\tau'_{\textnormal{I},n}}[1], \dots, x_{\textnormal{I},\tau'_{\textnormal{I},n}}[N - 1]\right]},
    \label{eq:vec_intf_sig}
\end{align} 
and $\tau'_{\textnormal{S},n} = \tau_{\textnormal{S},n} - \ell T_s$ and $\tau'_{\textnormal{I},n} = \tau_{\textnormal{I},n} - \ell T_s$ denote the delay of the $n$th path of the synchronization signal and the $n$th path of the interference signal offset by the start time of the current observation window, respectively.

To make the dependence of the received signal on $\tau_{\textnormal{S},n}$ explicit, we further transform $\bY_\ell$ into the DFT domain. The  matrix $\tilde{\bY}_\ell$ is obtained by applying the DFT to $\bY_\ell$ row-wise, i.e., $\tilde{\bY}_\ell = \bY_\ell\tp{\bF}$, and can be expressed as
\begin{align}
    \tilde{\bY}_{\ell} =\,& \sum_{n=1}^{L_\textnormal{S}} \bma_{\textnormal{S},n}\beta_{\textnormal{S},n}\tp{\tilde{\bmx}_{\textnormal{S},\tau'_{\textnormal{S},n}}} + \sum_{n=1}^{L_\textnormal{I}} \bma_{\textnormal{I},n}\beta_{\textnormal{I},n}\tp{\tilde{\bmx}_{\textnormal{I},\tau'_{\textnormal{I},n}}} + \tilde{\bZ} \nonumber \nonumber \\
     \mathrel{\overset{(1)}{=}} \, & \sum_{n=1}^{L_\textnormal{S}} \bma_{\textnormal{S},n} \beta_{\textnormal{S},n} \underbrace{\tp{\left(\bF\bmx_{\textnormal{S},0} \odot\bm\phi(\tau'_{\textnormal{S},n})\right)}}_{=\,\tp{\tilde{\bmx}}_{\textnormal{S}}\left(\tau'_{\textnormal{S}, n}\right)} \nonumber \\
    &  + \sum_{n=1}^{L_\textnormal{I}} \bma_{\textnormal{I},n}\beta_{\textnormal{I},n}\tp{\tilde{\bmx}_{\textnormal{I},\tau'_{\textnormal{I},n}}} + \tilde{\bZ},
    \label{eq:win_sig_model_freq}
\end{align}
where $\tilde{\bmx}_{\textnormal{S},\tau'_{\textnormal{S},n}} = \bF\bmx_{\textnormal{S},\tau'_{\textnormal{S},n}}$, $\tilde{\bmx}_{\textnormal{I},\tau'_{\textnormal{I},n}} = \bF\bmx_{\textnormal{I},\tau'_{\textnormal{I},n}}$, and $\bm\phi(\tau'_{\textnormal{S},n})$ is the linear phase shift given by
\begin{align}
\bm\phi(\tau'_{\textnormal{S},n}) &= \left[e^{-j 2\pi \omega_0 },e^{-j 2\pi \omega_1},\dots,e^{-j 2\pi \omega_{N-1}}\right], \nonumber \\
\omega_m &= \frac{k_m}{NT_s}\tau'_{\textnormal{S},n}
\label{eq:phase_ramp_fine}
\end{align}
with frequency indices $k_m$ following the DFT ordering:
\begin{align}
k_m =
\begin{cases}
m, & m = 0, 1, \dots, \lfloor (N-1)/2 \rfloor, \\
m - N, & m = \lceil N/2 \rceil, \dots, N-1.
\end{cases}
\label{eq:k_mapping_fine}
\end{align}

Given that the DFT is unitary, $\tilde{\bZ}$ follows the same distribution as $\bZ$. Step (1) applies the DFT time-shifting property. This is valid provided that the circular shift implied by the phase ramp coincides with the physical linear delay of the signal. The condition $N_\textnormal{P} > \ceil{\max_{n}\tau'_{\textnormal{S},n}/T_s}$ ensures that the synchronization signal's components do not wrap around the observation window.\footnote{For ToA estimation, it is only required that this condition holds for the first MPC delay---this is implicitly assumed for the rest of the paper.} In subsequent derivations, we use $\tilde{\bmx}_{\textnormal{S}}\left(\tau'\right)$, as defined in \fref{eq:win_sig_model_freq}, as a shorthand for the frequency-domain synchronization signal template at a delay $\tau'$.

\section{ToA Estimation Algorithm}

\subsection{General Formulation}
\label{sec:general_formulation}

Our goal is to estimate the delay of the first-arriving MPC of the synchronization signal, which we assume to be the ToA of interest, while mitigating the impact of all other propagation paths---be it from the same transmitter through multipath or a potential interferer. 
Equivalently, given that $\tau_{\textnormal{S},1} = \ell_{\textnormal{S},1}T_s + \tau'_{\textnormal{S},1}$, we seek to determine $\ell_{\textnormal{S},1} = \floor{\tau_{\textnormal{S},1}/T_s}$ and the delay offset~$\tau'_{\textnormal{S},1}$.

Our algorithm leverages the fact that all but one of the synchronization signal components in \fref{eq:win_sig_model_freq} can be treated as interference. The proposed method builds upon the spatial filtering idea put forward in \cite{Marti2025JammerResilience} for jammer mitigation, extending it to frequency-selective channels and subsample ToA estimation. We start by formulating two hypotheses:
\begin{align}
    \mathrm{H}_1&: \tilde{\bY}_{\ell} = \bmb_1 \tp{\tilde{\bmx}}_{\textnormal{S}}\left(\tau'_1\right) + \bB_\textnormal{I}\tp{\bX_\textnormal{I}} + \tilde{\bN}, \\
    \mathrm{H}_0&: \tilde{\bY}_{\ell} = \bB_\textnormal{I}\tp{\bX_\textnormal{I}} + \tilde{\bN}. 
    \label{eq:toa_hypothesis_test}
\end{align}
Hypothesis $\mathrm{H}_1$ assumes that the synchronization signal is present in the current window $\tilde{\bY}_\ell$ at a hypothesized delay~$\tau'_1$ with unknown spatial signature and gain modeled by $\bmb_1$; hypothesis $\mathrm{H}_0$ assumes that the synchronization sequence is absent and only nuisance MPCs are present.
The matrix~$\bB_\textnormal{I}$ aggregates the spatial signatures and gains of all nuisance propagation paths of the synchronization signal and all propagation paths of the true interference, while $\bX_\textnormal{I}$ aggregates the corresponding waveforms. To be precise, denoting $\bmb_{\textnormal{S},n} = \bma_{\textnormal{S},n}\beta_{\mathrm{S,n}}$ and $\bmb_{\textnormal{I},n} = \bma_{\textnormal{I},n}\beta_{\mathrm{I,n}}$, we have
\begin{align}
    \bB_\textnormal{I} &= \left[\left[\bmb_{\textnormal{S},n}\right]_{\bmb_{\textnormal{S},n}\neq\bmb_1}, \left[\bmb_{\textnormal{I},n}\right]\right] \\
    \bX_\textnormal{I} &= \left[\left[\tilde{\bmx}_\textnormal{S}\left(\tau'_{\textnormal{S},n}\right)\right]_{\tau'_{\textnormal{S},n}\neq\tau'_1}, \left[\tilde{\bmx}_{\textnormal{I},\tau'_{\textnormal{I},n}}\right]\right]
    \label{intf_mat_def}
\end{align}

It follows that $\rank\!\left(\bB_{\textnormal{I}}\tp{\bX}_\textnormal{I}\right) \leq L_\textnormal{S} + L_\textnormal{I}$, with equality when all delays and array response vectors are distinct for both the synchronization and interference MPCs and $\tau'_{\textnormal{S},n}\neq\tau'_1$. For the following derivations, we denote $I=\rank\!\left(\bB_{\textnormal{I}}\tp{\bX}_\textnormal{I}\right)$.

Based on the hypotheses $\mathrm{H}_1$ and $\mathrm{H}_0$, we employ a GLRT to derive a score function $\mathcal{S}(\ell, \tau'_1)$ (see \fref{sec:glrt_score_function}) that quantifies how well $\mathrm{H}_1$ explains the observed signal relative to $\mathrm{H}_0$. 
At the true delays of the synchronization signal MPCs, this function is expected to attain local maxima, i.e., $\mathrm{H}_1$ should better explain the received signal at those delays than in their immediate vicinity. 
Hence, to estimate the ToA of the received signal, we can move over consecutive receive blocks $\tilde{\bY}_\ell, \ell \in \opN_0$, and for each of these windows, sweep $\tau'_1 \in \left[0, T_s\right]$, stopping at the first local maximum that exceeds a given detection threshold~$\gamma$. Mathematically, this corresponds to solving
\begin{align}
    \{\hat{\ell}_{\mathrm{S},1}, \hat{\tau}'_{\mathrm{S},1}\}
    &=
    \min^{\textnormal{lex}}_{\substack{\ell \in \mathbb{N}_0,\, \tau'_1 \in [0,T_s] \\[2pt]
    \mathcal{S}(\ell,\tau'_1)\ge \gamma,\;
    \mathcal{S}'(\ell,\tau'_1)=0,\;
    \mathcal{S}''(\ell,\tau'_1)<0}}
    (\ell, \tau'_1),
    \label{eq:general_toa_problem}
\end{align}
where $\mathrm{lex}$ stands for \emph{lexicographic order}, which implies that $\ell$ takes precedence over $\tau'_1$, i.e., among all pairs $(\ell, \tau'_1)$ satisfying the constraints, the smallest $\ell$ is selected first, and for this~$\ell$, the smallest $\tau'_1$ is then chosen. 
We emphasize that our proposed ToA estimation algorithm follows this approach, with a few simplifying approximations that reduce complexity. The resulting procedure is summarized next.

\subsection{GLRT Score Function}
\label{sec:glrt_score_function}

We first derive the score function $\mathcal{S}(\ell, \tau'_1)$. To this end, we introduce the projector onto $\mathrm{span}\left(\tp{\tilde{\bmx}_\textnormal{S}}\left(\tau'_{1}\right)\right)$, defined as
\begin{align}
\tilde{\mathbf{\Pi}}_\textnormal{S}\left(\tau'_{1}\right) &= \frac{\conj{\tilde{\bmx}_\textnormal{S}}\left(\tau'_{1}\right)}{\vecnorm{\tilde{\bmx}_\textnormal{S}\left(\tau'_{1}\right)}}\frac{\tp{\tilde{\bmx}_\textnormal{S}}\left(\tau'_{1}\right)}{\vecnorm{\tilde{\bmx}_\textnormal{S}\left(\tau'_{1}\right)}} = \tilde{\bmu}_\textnormal{S}\left(\tau'_{1}\right) \herm{\tilde{\bmu}_\textnormal{S}}\left(\tau'_{1}\right),
    \label{eq:proj_s}
\end{align}
and the projector onto its orthogonal complement, defined as 
\begin{align}
    \tilde{\bT}\left(\tau'_{1}\right) = \matI_N - \tilde{\mathbf{\Pi}}_\textnormal{S}\left(\tau'_{1}\right).
    \label{eq:proj_orth_s}
\end{align} 

\subsubsection{Hypothesis $\mathrm{H}_1$}
Starting with $\mathrm{H}_1$, the log-likelihood of the unknown parameters can be expressed as
\begin{align}   &\ell_1\!\left(\bmb_1,\bB_\textnormal{I},\bX_\textnormal{I},\No \vert \tilde{\bY}_\ell \right) = - MN \ln\left(\pi \No\right) \nonumber \\
    & \qquad \qquad \qquad - \frac{1}{\No}\underbrace{\frobnorm{\tilde{\bY}_\ell - \bmb_1 \tp{\tilde{\bmx}}_{\textnormal{S}}\left(\tau'_{1}\right) - \bB_\textnormal{I}\tp{\bX_\textnormal{I}}}^2}_{=\,Q_1}.
    \label{eq:h1_log_likelihood}
\end{align}

The quantity $Q_1$ can be further decomposed by exploiting the orthogonality property of the Frobenius norm, yielding
\begin{align}
    Q_1 = \, & \frobnorm{\tilde{\bY}_\ell\tilde{\bT}\left(\tau'_{1}\right) - \bB_\textnormal{I}\tp{\bX_\textnormal{I}}\tilde{\bT}\left(\tau'_{1}\right)}^2 \nonumber \\ 
    & + \frobnorm{\tilde{\bY}_\ell\tilde{\mathbf{\Pi}}_\textnormal{S}\left(\tau'_{1}\right) - \bmb_1 \tp{\tilde{\bmx}}_{\textnormal{S}}\left(\tau'_{1}\right) - \bB_\textnormal{I}\tp{\bX_\textnormal{I}}\tilde{\mathbf{\Pi}}_\textnormal{S}\left(\tau'_{1}\right)}^2 \nonumber \\
    =\, & \frobnorm{\tilde{\bY}_{\ell, \orth}\left(\tau'_{1}\right) - \bU_\textnormal{I} \bX'}^2 \! + \vecnorm{\bmz_{\ell}\left(\tau'_{1}\right) - \bmb'_1 - \bU_\textnormal{I} \bmc}^2\!.
    \label{eq:frob_norm1_decomposition}
\end{align}
Let $\bU_\textnormal{I}\mathbf\Sigma_\textnormal{I}\herm{\bV}_\textnormal{I}$ be the SVD of $\bB_\textnormal{I}$. Then $\tilde{\bY}_{\ell, \orth}\left(\tau'_{1}\right) = \tilde{\bY}_\ell\tilde{\bT}\left(\tau'_{1}\right)$, $\bmz_{\ell}\left(\tau'_{1}\right) = \tilde{\bY}_\ell\tilde{\bmu}_\textnormal{S}\left(\tau'_{1}\right)$, $\bmb'_1 = \bmb_1 \vecnorm{{\tilde{\bmx}}_{\textnormal{S}}\left(\tau'_{1}\right)}$, $\bX' = \mathbf\Sigma_\textnormal{I}\herm{\bV}_\textnormal{I}\tp{\bX}_\textnormal{I}\tilde{\bT}\left(\tau'_{1}\right)$ and $\bmc = \mathbf\Sigma_\textnormal{I}\herm{\bV}_\textnormal{I}\tp{\bX}_\textnormal{I}\tilde{\bmu}_\textnormal{S}\left(\tau'_{1}\right)$.
The unknown parameters $\bU_\textnormal{I}$, $\bX'$, $\bmb'_1$, and $\bmc$ can be estimated via maximum likelihood estimation by minimizing $Q_1$. The minimization of the second term yields the trivial least-squares solution $\est{\bmb}'_1=\bmz_{\ell} \left(\tau'_1\right) - \bU_\textnormal{I} \bmc$, which, when substituted back into $Q_1$, collapses the second term to zero. Hence, we obtain
\begin{align}
    \{\est{\bU}_\textnormal{I}, \est{\bX}'\} = \mathrm{arg}\,\min_{\bU_\textnormal{I}, \bX'} \,\frobnorm{\tilde{\bY}_{\ell, \orth}\left(\tau'_{1}\right) - \bU_\textnormal{I} \bX'}^2.
    \label{eq:h1_ll_min}
\end{align}

For $\bX'$, this corresponds to the least-squares solution
\begin{align}
    \est{\bX}' = \herm{\bU}_{\textnormal{I}}\tilde{\bY}_{\ell, \orth}\left(\tau'_{1}\right),
    \label{eq:h1_B_est}
\end{align}
and substituting it back into \fref{eq:h1_ll_min}, we obtain:
\begin{align}
    \est{\bU}_\textnormal{I} = \mathrm{\arg} \, \min_{\bU_\textnormal{I}} \, \frobnorm{\tilde{\bY}_{\ell, \orth}\left(\tau'_{1}\right) - \bU_\textnormal{I}\herm{\bU}_{\textnormal{I}}\tilde{\bY}_{\ell, \orth}\left(\tau'_{1}\right)}^2.
    \label{eq:h1_profiled_ll_min}
\end{align}

According to the Eckart-Young-Mirsky theorem \cite{Eckart1936Approximation}, $\est{\bU}_\textnormal{I}$ is given by the leading $I$ singular vectors of $\tilde{\bY}_{\ell, \orth}\left(\tau'_{1}\right)$ as
\begin{align}
    \est{\bU}_\textnormal{I} = \bU_{\tilde{\bY}_{\ell, \orth}\left(\tau'_{1}\right),1:I}.
    \label{eq:h1_U_est}
\end{align}

Substituting back all estimates into $Q_1$ results in
\begin{align}
    Q_1\!\left(\est{\bU}_\textnormal{I}, \est{\bX}', \est{\bmb}'_1, \bmc\right) &=\frobnorm{\tilde{\bY}_{\ell, \orth} \left(\tau'_{1}\right) - \est{\bU}_\textnormal{I}\est{\bX}'}^2 \nonumber \\
    &= \Bigl\|\underbrace{\left(\matI_M - \est{\bU}_\textnormal{I}\herm{\est{\bU}}_\textnormal{I}\right)}_{=\,\tilde{\bP}_\ell\left(\tau'_{1}\right)}\tilde{\bY}_\ell\tilde{\bT}\left(\tau'_{1}\right)\Bigr\|^2_\textnormal{F},
    \label{eq:h1_residual}
\end{align}
where $\tilde{\bP}_\ell\left(\tau'_{1}\right)$ is the projector onto the orthogonal complement of the estimated interference subspace at delay $\tau'_{1}$. Finally, the unknown noise variance can be estimated as\footnote{$\hat{\mathbf P}(\tau'_1)$ and $\mathbf T(\tau'_1)$, with rank $M - I$ and $N - 1$, reduce the effective number of degrees of so the factor $MN$ is replaced by $(M - I)(N - 1)$.}
\begin{align}
    \est{N}_0 &= \mathrm{arg} \, \max_{\No}\bigg[-(M-I)(N-1) \ln\left(\pi \No\right)  \nonumber \\
    & \qquad \qquad  \qquad \! \! - \frac{1}{\No}\frobnorm{\tilde{\bP}_\ell\left(\tau'_{1}\right)\tilde{\bY}_{\ell} \tilde{\bT}\left(\tau'_{1}\right)}^2\bigg] \\
    &= \frobnorm{\tilde{\bP}_\ell\left(\tau'_{1}\right)\tilde{\bY}_{\ell} \tilde{\bT}\left(\tau'_{1}\right)}^2/(M-I)(N-1).
    \label{eq:n0_est}
\end{align}

\subsubsection{Hypothesis $\mathrm{H}_2$}
Moving to $\mathrm{H}_0$, the log-likelihood of the unknown parameters can be expressed as
\begin{align}
    \ell_0\left(\bB_\textnormal{I},\bX_\textnormal{I},\No \vert \tilde{\bY}_\ell \right) = \,& - MN \ln\left(\pi \No\right) \nonumber \\
    & - \frac{1}{\No}\underbrace{\frobnorm{\tilde{\bY}_\ell - \bB_\textnormal{I}\tp{\bX_\textnormal{I}}}^2}_{=\,Q_0}.
    \label{eq:h0_log_likelihood}
\end{align}

Analogous to \fref{eq:frob_norm1_decomposition}, the quantity $Q_0$ can be decomposed as
\begin{align}
    Q_0 &= \frobnorm{\tilde{\bY}_{\ell, \orth}\left(\tau'_{1}\right) - \bU_\textnormal{I} \bX'}^2 + \frobnorm{\bmz_{\ell}\left(\tau'_{1}\right) - \bU_\textnormal{I} \bmc}^2.
    \label{eq:frob_norm0_decomposition}
\end{align}

In this case, the optimal estimate of $\mathbf{U}_\textnormal{I}$ is not strictly determined by the first term of \fref{eq:frob_norm0_decomposition}. Nevertheless, the projection $\tilde{\bT}(\tau'_{1})$ only attenuates the nuisance MPCs---reducing the effective interference-to-noise ratio (INR) without altering the column space of the received signal matrix. Thus, the singular vectors estimated from the projected signal $\tilde{\mathbf{Y}}_{\ell, \perp\left(\tau'_{1}\right)}$ and from the full observation $\tilde{\mathbf{Y}}_\ell$ coincide when $\textnormal{INR}\to\infty$ or $N\to\infty$. 
Therefore, to reduce complexity and forego the need for a second subspace estimate, we reuse the estimate $\est{\bU}_\textnormal{I}$ from \fref{eq:h1_U_est} for $\mathrm{H}_0$. As a further approximation, we also reuse the noise variance estimate $\est{N}_0$ from \eqref{eq:n0_est} for $\mathrm{H}_0$.\footnote{A separate estimate could be obtained from $\tilde{\bP}_\ell(\tau'_1)\tilde{\bY}_\ell$ with slightly lower variance. However, since the noise model is identical under both hypotheses and the additional projection $\tilde{\bT}(\tau'_1)$ does not affect the noise power, reusing~$\est{N}_0$ remains (asymptotically) unbiased and statistically consistent.}

With these two approximations, the score function becomes
\begin{align}
    \mathcal{S}\left(\ell,\tau'_{1}\right) = \, & \max_{\bX',\bmb'_1, \bmc} -MN \ln\left(\est{N}_0\right)-\frac{1}{\est{N}_0}Q_1\left(\est{\bU}_\textnormal{I},\bX',\bmb'_1,\bmc\right) \nonumber \\
    & -\max_{\bX', \bmc} -MN \ln\left(\est{N}_0\right) -\frac{1}{\est{N}_0}Q_0\left(\est{\bU}_\textnormal{I},\bX',\bmc\right) \nonumber \\
    =\,& -\max_{\bmc} -\frac{1}{\est{N}_0} \vecnorm{\bmz_{\ell}\left(\tau'_{1}\right) - \est{\bU}_\mathrm{J} \bmc}^2 \nonumber \\
    = \, &\frac{\vecnorm{\tilde{\bP}_\ell\left(\tau'_{1}\right)\tilde{\bY}_{\ell}\tilde{\bmu}_\textnormal{S}\left(\tau'_{1}\right)}^2 }{\frobnorm{\tilde{\bP}_\ell\left(\tau'_{1}\right)\tilde{\bY}_{\ell}\tilde{\bT}\left(\tau'_{1}\right)}^2} (M-I)(N-1).
    \label{eq:glrt_test_stat}
\end{align}

This score function has an intuitive interpretation: it first applies a spatial filter to suppress all components except the one at the hypothesized delay, then correlates the filtered output with the corresponding reference waveform, normalized by its energy. The score function is inversely weighted by the estimated noise power, reducing confidence in the hypothesized component as the noise power increases.

\subsection{Approximate Procedure}
\label{sec:approximate_solution}

Solving the ToA estimation problem as stated in \fref{eq:general_toa_problem} would result in prohibitive complexity: for every new signal sample and every single delay $\tau'_1$ tested within the corresponding window---in principle over a very fine grid---a new partial eigenvalue decompostion (EVD) of the $M$-by-$M$ covariance matrix of $\tilde{\bY}_{\ell, \orth}\left(\tau'\right)$ would have to be performed to determine the leading $I$ singular vectors required to build $\tilde{\bP}_\ell(\tau')$. 
To reduce complexity, we propose to split the optimization problem into two sequential stages: (i) coarse ToA estimation and (ii) fine ToA estimation.

\subsubsection{Coarse ToA Estimation}
\label{sec:coarse_toa_estimation}

Coarse ToA estimation determines only the sample index $\ell_{\textnormal{S},1}$. For this stage, we fix $\tau'_{1} = 0$ and assume a rank-$I_\textnormal{C}$ approximation of the interference subspace. Because $\tau'_{1}$ is fixed, transforming the received signal to the frequency-domain becomes unnecessary, so the coarse ToA problem is formulated as
\begin{align}
    \est{\ell}_{\textnormal{S}, 1} &= \min_{\ell\in\opN_0} \; \ell\quad \text{s.t.}\quad \frac{\vecnorm{\bP_\ell\bY_{\ell}\bmu_\textnormal{S}}^2 }{\frobnorm{\bP_\ell\bY_{\ell}\bT}^2} D > \gamma',
    \label{eq:coarse_toa_opt}
\end{align}
where $\bmu_\textnormal{S} = \conj{\bmx}_{\textnormal{S},0}/ \vecnorm{\bmx_{\textnormal{S},0}}$, $\bT=\matI_{N}-\bmu_\textnormal{S}\herm{\bmu_\textnormal{S}}$, $\bP_\ell = \matI_{M} - \est{\bU}_{\textnormal{I}}\herm{\est{\bU}}_{\textnormal{I}}$, with $\est{\bU}_\textnormal{I}=\bU_{\bY_\ell,1:I_\textnormal{C}}$, and $D=(M-I_\textnormal{C})(N-1)$. This stage is deemed successful if $\est{\ell}_{\textnormal{S},1} \in \left\{\ell_{\textnormal{S},1}-1,\ell_{\textnormal{S},1},\ell_{\textnormal{S},1}+1\right\}$. We allow for an error of $\pm 1$ sample to account for the fact that $\bmu_\textnormal{S}$ will not be perfectly aligned with the true $\bmu_{\textnormal{S},\tau'_{\textnormal{S},1}}=\conj{\bmx}_{\textnormal{S},\tau'_{\textnormal{S},1}}/ \big\|\bmx_{\textnormal{S},\tau'_{\textnormal{S},1}}\big\|$. In the fine ToA estimation stage, we then sweep the subsample delay $\tau'_{\textnormal{S},1}$ over an extended interval $\left[0,2T_s\right]$ to account for the sample index uncertainty.

\subsubsection{Fine ToA Estimation}
\label{sec:fine_toa_estimation}
Fine ToA estimation determines only the subsample offset $\tau'_{\textnormal{S}, 1}$ within the window set by the previous stage. We fix $\ell=\est{\ell}_{\textnormal{S},1} - 1$ and assume a rank-$I_\textnormal{F}$ approximation of the interference subspace, with $I_\textnormal{C} \leq I_\textnormal{F} \leq M-2$. We formulate the fine ToA estimation problem as
\begin{align}
\hat{\tau}'_{\mathrm{S},1}
&=
\operatorname*{min}_{\substack{
\tau'_1 \in [0,2T_s] \\
\mathcal{S}'(\hat{\ell}_{\mathrm{S},1}-1,\tau'_1)=0,\;
\mathcal{S}''(\hat{\ell}_{\mathrm{S},1}-1,\tau'_1)<0}}
\tau'_1,
\label{eq:fine_toa_opt}
\end{align}
that is, we set $\est{\tau}'_{\textnormal{S},1}$ as the first local maximum of the score function evaluated for $\tilde{\bY}_{\est{\ell}_{\textnormal{S},1}-1}$. 
To find this maximum efficiently, we partition the search interval $[0,\, 2T_s]$ into $J$ sub-intervals of length $h = 2T_s/J$, with endpoints $\tau'_{\textnormal{E},j} = jh$, $j=0,\dots,J-1$ and find the smallest indices $j^\ast$, $j^\ast+1$ for which the derivative of the score function undergoes a sign change from positive to negative. Formally,
\begin{align}
j^\ast
&=
\min_{\substack{
j \in \{0,\dots,J-2\} \\
\mathcal{S}'(\hat{\ell}_{\mathrm{S},1}-1,\tau'_{\textnormal{E},j}) > 0,\;
\mathcal{S}'(\hat{\ell}_{\mathrm{S},1}-1,\tau'_{\textnormal{E},j+1}) < 0}} j
\label{eq:sub_interval_search_fine}
\end{align}

This procedure brackets the first local maximum inside the $[\tau'_{\textnormal{E},j^\ast}, \tau'_{\textnormal{E},j^\ast+1}]$ interval, significantly reducing the search space for $\est{\tau}'_{\textnormal{S},1}$. 
To compute the derivative of the score function, $\mathcal{S}'(\hat{\ell}_{\mathrm{S},1}-1,\tau'_1)$, we implement both estimation stages in JAX \cite{Bradbury2018JAX} and leverage its automatic differentiation capabilities, circumventing the need for deriving a closed-form expression of the derivative. 
In addition, JAX also allows for just-in-time compilation, further accelerating the evaluation of both the score function and its derivative. 

Once a bracket is found, we employ the golden-section search (GSS)~\cite{Press2007GoldenSection} method for refining the estimate. Denoting $\mathrm{GSS}(f\left(\cdot\right),a,b)$ as the golden-section search applied to $f\left(\cdot\right)$ on the $[a,b]$ interval, the final estimate is given by
\begin{align}
\est{\tau}'_{\textnormal{S},1} = \mathrm{GSS}\!\left( \mathcal{S}(\est{\ell}_{\textnormal{S},1}-1,\cdot), \tau'_{\textnormal{E},j^\ast}, \tau'_{\textnormal{E},j^\ast+1} \right).
\label{eq:gss}
\end{align} 

This two-step procedure substantially reduces the number of costly score function evaluations and accelerates ToA estimation. For it to be valid, however, $J$ must be sufficiently large to ensure that each sub-interval contains at most one maximum, satisfying the unimodality assumption of the GSS method. Equivalently, the sub-interval length $h$ should be set to the minimum resolvable delay separation, $\Delta \tau'_\text{min}$, and $J \geq \ceil{2/\Delta\tau'_{\text{min}}}$. Below this limit, the score function exhibits merged minima, and a finer partitioning offers no advantage.  

\subsubsection{Subsample Delay Interval Partitioning}
\label{sec:delay_intv_part}
To complete the fine ToA estimation procedure, we need to determine a suitable partitioning for the subsample delay interval, or equivalently, an upper bound on $J$, denoted by $J_\textnormal{max}$.\footnote{The resulting upper bounds are specific only to the tested synchronization signal. They remain valid for any other environment and system parameters.} 
To that end, for various SNR values, we consider a worst-case scenario with two propagation paths with unitary gain and a separation of $90^\circ$ in azimuth and elevation, setting the synchronization signal to the one later used for global performance evaluation---a Zadoff--Chu sequence of length $K = 63$, filtered by a RRC filter spanning $R = 9$ symbols with a roll-off factor of $0.3$, oversampled by a factor of $P = 2$. We also consider the highest rank approximation that is later tested ($I_\textnormal{F} = 16$). 
To determine delay resolvability, we sweep the separation $\Delta\tau'= \abs{\tau'_{\textnormal{S}, 1} - \tau'_{\textnormal{S},2}}$ in the interval $\left[0,2T_s\right]$ and perform a Monte--Carlo simulation with $20'000$ trials for each evaluated point. For every point, we compute the ratio of trials for which separation is successful, i.e., when exactly two derivative zero-crossings can be detected in the proposed score function. In~\fref{tbl:sub_intv_num}, we summarize the delay separations obtained for a success rate $r_\textnormal{S}$ of $90\%$, $50\%$, and $10\%$. To be conservative, we take $\Delta \tau'_\text{min} = \Delta\tau'_{10\%}$ for each SNR. The corresponding number of sub-intervals $J_\text{max}$ is used in every following experiment.

\begin{table}[tbp]
\centering
\caption{Summary of delay resolvability.}
\label{tbl:sub_intv_num}	
\renewcommand{\arraystretch}{1.1}

\begin{tabular}{@{}lc*{3}{c}@{}}
\toprule
\multirow{3}{*}{SNR} & \multicolumn{3}{c}{Delay Separation $\Delta\tau'_{r_\textnormal{S}} [\mathrm{Samples}]$} & $J_\text{max}$ \\
\cmidrule(lr){2-4}

& $\Delta\tau'_{90\%}$ & $\Delta\tau'_{50\%}$ & $\Delta\tau'_{10\%}$ & \\ 
\midrule

$10\,\mathrm{dB}$ & $0.32$ & $0.29$ & $0.24$ & $5$ \\ 
$20\,\mathrm{dB}$ & $0.18$ & $0.16$ & $0.14$ & $8$ \\ 
$30\,\mathrm{dB}$ & $0.10$ & $0.09$ & $0.08$ & $13$ \\ 

\bottomrule
\end{tabular}
\end{table}

As expected, delay resolvability improves with increasing SNR. In any case, super-resolution is achieved---considering the delay separation at the $90\%$ success rate point, we observe a three-fold resolution improvement for an SNR of $10\,\text{dB}$ and a ten-fold improvement for an SNR of $30\,\text{dB}$ with respect to the resolution that would be achieved with standard correlation.
\section{Results}
\label{sec:results}

\subsection{Simulated Scenario}
\label{sec:simulated_scenario}

\begin{figure}[tp]
    \centering
    \definecolor{activetransmitter}{HTML}{00FF00} 
    \definecolor{transmittercolor}{HTML}{0000FF} 
    \definecolor{receivercolor}{HTML}{750402}
    \definecolor{activereceivercolor}{HTML}{EDB120}
    
    \begin{tikzpicture}
        \matrix[
            column sep=0.2cm,
            row sep=0.0cm,
            draw, fill=white,
            inner sep=0.05cm,
            nodes={align=left, anchor=west}
        ] {
            \node {\tikz{\fill[transmittercolor] (0,0) circle (0.06);
                        \node[anchor=west] at (0.15,0) {Receiver Locations};}}; &
            \node {\tikz{\fill[activetransmitter] (0,0) circle (0.06);
                        \node[anchor=west] at (0.15,0) {Active Receiver};}}; \\
            \node {\tikz{\fill[receivercolor] (0,0) circle (0.06);
                        \node[anchor=west] at (0.15,0) {Transmitter Locations};}}; &
            \node {\tikz{\fill[activereceivercolor] (0,0) circle (0.06);
                        \node[anchor=west] at (0.15,0) {TXs in RX LoS FOV};}}; \\
        };
    \end{tikzpicture}
    
    \subfloat[Detailed 3D view.\label{fig:full_3d_model}]{%
        \includegraphics[width=0.5\linewidth]{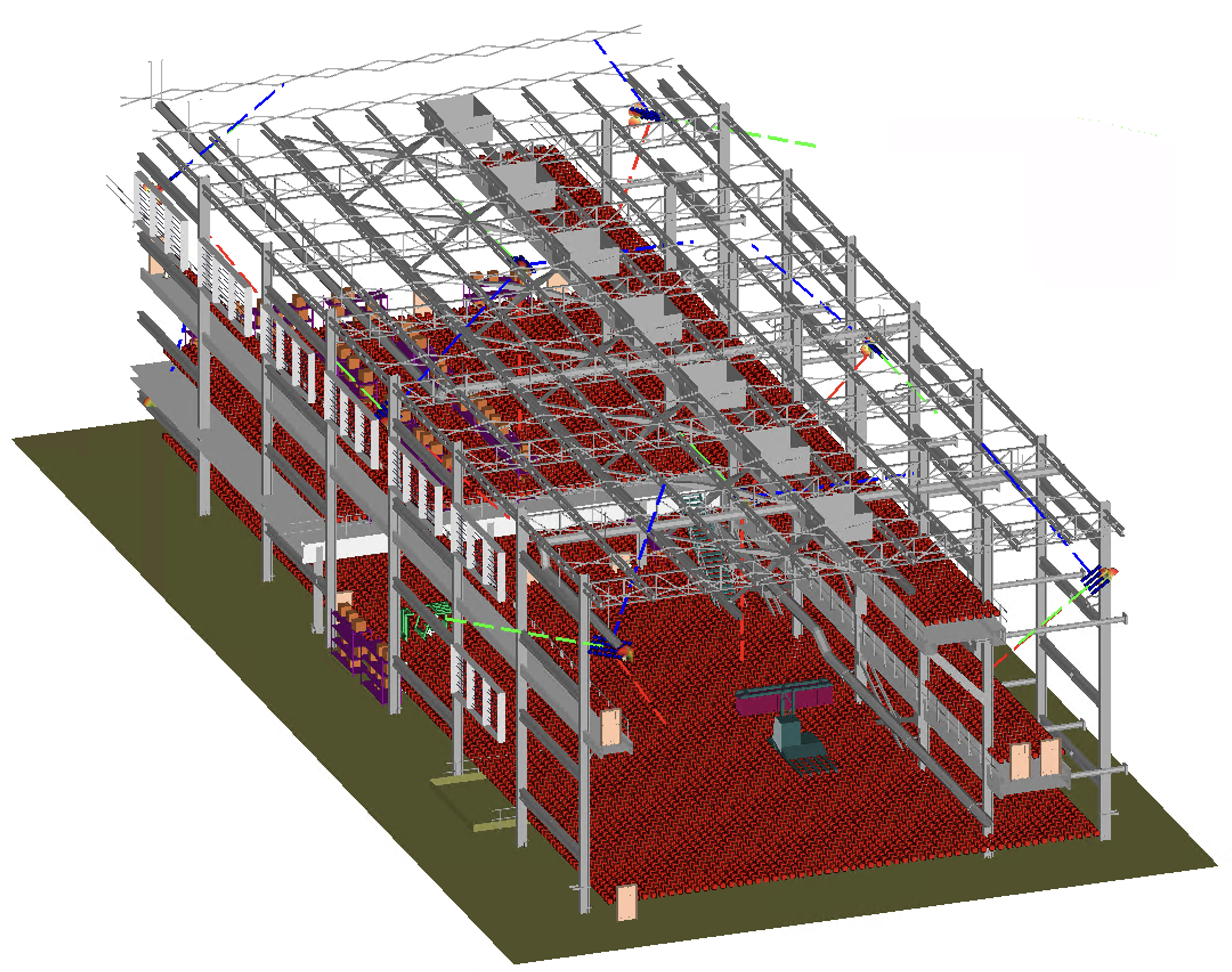}%
    }%
    \subfloat[Simplified 3D view.\label{fig:simplified_3d_model}]{%
        \includegraphics[width=0.5\linewidth]{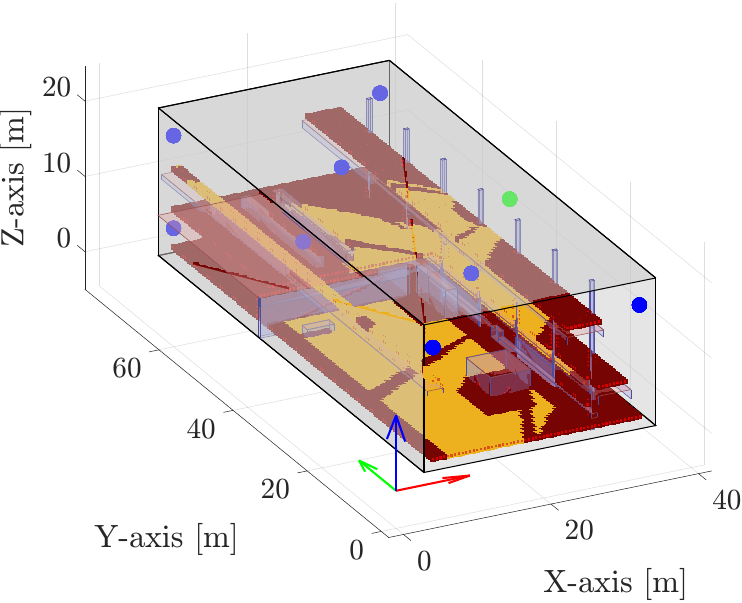}%
    }
    \caption{Simulated indoor factory propagation scenario. (a) depicts the view from Wireless InSite and (b) the reconstructed view in MATLAB. (b) shows walls, floors, and simplified bounding boxes around the most prominent objects in the environment, as well as the transmitters in line-of-sight of one of the wall-mounted receivers, accounting for a limited $120^\circ$ FOV and occlusions.}
    \label{fig:test_scenario}
\end{figure}
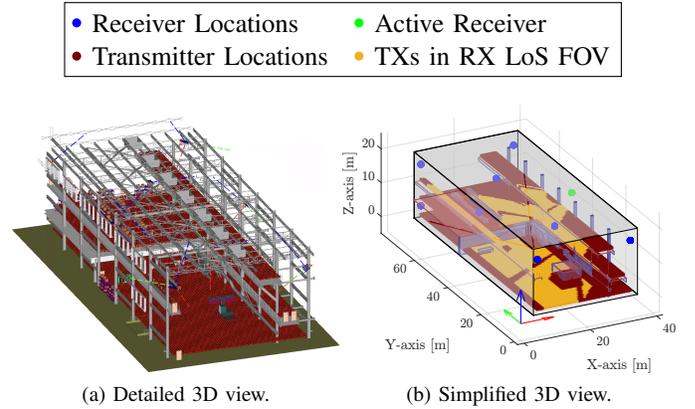

We evaluate the proposed algorithm using ray-traced channel impulse responses (CIRs) for an indoor factory scenario (\fref{fig:test_scenario}) generated with Remcom’s Wireless InSite~\cite{Remcom2025WirelessInSite}.\footnote{The CIR dataset is available at \url{https://zenodo.org/records/17566853}} CIRs are selected between random pairs of omnidirectional transmitters (one legitimate, one interferer) placed $1.2$\,m above each factory floor in a rectangular grid, and receivers with $8\times4$ URAs. Transmitter locations are restricted to those with line-of-sight to the receiver, within a $120^\circ$ cone centered on the array's boresight. We consider two distinct Zadoff-Chu sequences of length $K = 63$ for the synchronization sequence and interference signals, a carrier frequency of $15\mathrm{GHz}$, an oversampling factor $P = 2$, and a root-raised-cosine pulse-shaping filter spanning $R = 9$ symbols with a roll-off factor of $0.3$.
The noise variance is given by $\No = k_\text{B}f_\text{s}T_\text{sys}10^{F_\text{NF}/10}$, where $T_\text{sys} = 298.15\,\text{K}$, $F_\text{NF} = 3\,\text{dB}$, and $k_\text{B}$ is the Boltzmann constant. Simulation results are obtained for a symbol rate of $R_s = 100\,\text{Msps}$ and a sampling frequency $f_\text{s} = 200\,\text{MHz}$. 
We evaluate signal-to-interference ratios (SIRs) from $-20\,\text{dB}$ to $20\,\text{dB}$, and various SNRs by dynamically adjusting the interference or synchronization signal transmit power. 

\subsection{Performance Metrics}

\subsubsection{Coarse ToA Estimation}
\label{sec:perf_metrics_coarse_toa}

Errors at this stage are false alarms ($\est{\ell}_{\textnormal{S},1} < \ell_{\textnormal{S},1}-1$) or missed detections ($\est{\ell}_{\textnormal{S},1} > \ell_{\textnormal{S},1}+1$). Commonly, performance is measured in terms of the receiver operating characteristic (ROC) curve, relating false alarm rate to missed detection rate by varying the detection threshold $\gamma'$ in \fref{eq:coarse_toa_opt}. 
However, to better compare different methods over a range of SIRs with a single scalar, we take the area under the ROC (AUC). An AUC of $0$ corresponds to a perfect detection method and an AUC of $0.5$ to random guessing. 

\subsubsection{Fine ToA Estimation}
The performance of this stage is evaluated with either the CDF or the median of the absolute timing error (measured in samples). CDFs compare the performance of different methods at fixed SIR; the median error tracks performance across SIR ranges in ablation studies.

\subsection{Simulation Results}
\label{sec:simulation_results}
\subsubsection{Coarse ToA Estimation}

\begin{figure}[tp]
    \centering
%
\definecolor{mycolor1}{rgb}{0.54902,0.65490,0.75686}%
\definecolor{mycolor3}{rgb}{0.54510,0.00000,0.00000}%
\definecolor{mycolor9}{rgb}{0.24314,0.41569,0.58039}%
\definecolor{mycolor10}{rgb}{0.04314,0.23922,0.56863}%

\begin{tikzpicture}

\begin{axis}[%
width=\linewidth,
height=0.618\linewidth,
xmin=-24,
xmax=33,
xtick={-20,-10,0,10,20,30},
xticklabels={{-20},{-10},{0},{10},{20},{$\infty$}},
xlabel style={font=\color{white!15!black}},
xlabel={SIR [dB]},
ymin=-0.05,
ymax=0.55,
ylabel style={font=\color{white!15!black}},
ylabel={AUC},
axis background/.style={fill=white},
axis x line*=bottom,
axis y line*=left,
xmajorgrids,
ymajorgrids,
legend style={legend cell align=left, align=left, draw=white!15!black}
]

\addplot [color=mycolor3, dashed, line width=2.0pt]
  table[row sep=crcr]{%
-20	0.488060235977173\\
-19	0.483892917633057\\
-18	0.477295368909836\\
-17	0.465568751096725\\
-16	0.44342577457428\\
-15	0.397852510213852\\
-14	0.315524518489838\\
-13	0.215109020471573\\
-12	0.146513909101486\\
-11	0.109296396374702\\
-10	0.0861181765794754\\
-9	0.0716355815529823\\
-8	0.0609907954931259\\
-7	0.0531904771924019\\
-6	0.0470699854195118\\
-5	0.0421237088739872\\
-4	0.0385320596396923\\
-3	0.0357025563716888\\
-2	0.0334746167063713\\
-1	0.0321086347103119\\
0	0.0315527245402336\\
1	0.0319658331573009\\
2	0.0325580537319183\\
3	0.0331496894359589\\
4	0.033315546810627\\
5	0.0332830920815468\\
6	0.0333070158958435\\
7	0.0333355143666267\\
8	0.0333664789795876\\
9	0.0333629250526428\\
10	0.0333299413323402\\
11	0.0333666987717152\\
12	0.0333700813353062\\
13	0.0333945266902447\\
14	0.0333719328045845\\
15	0.0333856642246246\\
16	0.0334027335047722\\
17	0.0333931036293507\\
18	0.0334079936146736\\
19	0.0334018692374229\\
20	0.0334112085402012\\
};
\addlegendentry{TDNC}

\addplot [color=mycolor3, only marks, mark=square*, mark options={solid, fill=mycolor3, draw=mycolor3}, forget plot]
  table[row sep=crcr]{%
-20	0.488060235977173\\
};
\node[fill=white, above, align=center, inner sep=0, font=\color{mycolor3}]
at (axis cs:-20,0.503) {0.488};
\addplot [color=mycolor3, only marks, mark=square*, mark options={solid, fill=mycolor3, draw=mycolor3}, forget plot]
  table[row sep=crcr]{%
-10	0.0861181765794754\\
};
\node[fill=white, above, align=center, inner sep=0, font=\color{mycolor3}]
at (axis cs:-10,0.101) {0.086};
\addplot [color=mycolor3, only marks, mark=square*, mark options={solid, fill=mycolor3, draw=mycolor3}, forget plot]
  table[row sep=crcr]{%
0	0.0315527245402336\\
};
\node[fill=white, above, align=center, inner sep=0, font=\color{mycolor3}]
at (axis cs:0,0.047) {0.032};
\addplot [color=mycolor3, only marks, mark=square*, mark options={solid, fill=mycolor3, draw=mycolor3}, forget plot]
  table[row sep=crcr]{%
10	0.0333299413323402\\
};
\node[fill=white, above, align=center, inner sep=0, font=\color{mycolor3}]
at (axis cs:10,0.048) {0.033};
\addplot [color=mycolor3, only marks, mark=square*, mark options={solid, fill=mycolor3, draw=mycolor3}, forget plot]
  table[row sep=crcr]{%
20	0.0334112085402012\\
};
\node[fill=white, above, align=center, inner sep=0, font=\color{mycolor3}]
at (axis cs:20,0.048) {0.033};
\addplot [color=mycolor3, only marks, mark=square*, mark options={solid, fill=mycolor3, draw=mycolor3}, forget plot]
  table[row sep=crcr]{%
30	0.0334097184240818\\
};
\node[fill=white, above, align=center, inner sep=0, font=\color{mycolor3}]
at (axis cs:30,0.048) {0.033};
\addplot [color=mycolor3, dotted, line width=2.0pt, forget plot]
  table[row sep=crcr]{%
20	0.0334112085402012\\
30	0.0334097184240818\\
};
\addplot [color=mycolor3, only marks, mark size=2.0pt, mark=square*, mark options={solid, fill=mycolor3, draw=mycolor3}, forget plot]
  table[row sep=crcr]{%
30	0.0334097184240818\\
};

\addplot [color=mycolor1, line width=2.0pt]
  table[row sep=crcr]{%
-20	0.111055359244347\\
-19	0.0952686369419098\\
-18	0.0811465233564377\\
-17	0.0692898333072662\\
-16	0.0602070614695549\\
-15	0.0535393506288528\\
-14	0.0485334917902946\\
-13	0.0445775836706161\\
-12	0.0415848977863789\\
-11	0.0393154546618462\\
-10	0.0376172512769699\\
-9	0.0363575667142868\\
-8	0.0353108607232571\\
-7	0.0345483645796776\\
-6	0.0338529534637928\\
-5	0.0331599824130535\\
-4	0.0324044115841389\\
-3	0.0313235260546207\\
-2	0.0296547096222639\\
-1	0.0270433034747839\\
0	0.0237079933285713\\
1	0.0209382548928261\\
2	0.0193409230560064\\
3	0.0181619580835104\\
4	0.0170714929699898\\
5	0.0160527396947145\\
6	0.0152076417580247\\
7	0.0144877713173628\\
8	0.0138193806633353\\
9	0.0133053474128246\\
10	0.0128988921642303\\
11	0.0125990696251392\\
12	0.0124363824725151\\
13	0.0123882424086332\\
14	0.0124044828116894\\
15	0.0124467713758349\\
16	0.0125518962740898\\
17	0.012708337046206\\
18	0.0128872264176607\\
19	0.0130638293921947\\
20	0.0132105369120836\\
};
\addlegendentry{Ours ($I_\textnormal{C}$=1)}

\addplot [color=mycolor1, dotted, line width=2.0pt, forget plot]
  table[row sep=crcr]{%
20	0.0132105369120836\\
30	0.014374959282577\\
};
\addplot [color=mycolor1, only marks, mark size=2.0pt, mark=*, mark options={solid, fill=mycolor1, draw=mycolor1}, forget plot]
  table[row sep=crcr]{%
30	0.014374959282577\\
};

\addplot [color=mycolor9, line width=2.0pt]
  table[row sep=crcr]{%
-20	0.0439757704734802\\
-19	0.0401270017027855\\
-18	0.0370562300086021\\
-17	0.0345312654972076\\
-16	0.0323314890265465\\
-15	0.0304267965257168\\
-14	0.0286775901913643\\
-13	0.0271726101636887\\
-12	0.0258280355483294\\
-11	0.0243836492300034\\
-10	0.0230101384222507\\
-9	0.0217077806591988\\
-8	0.0205384306609631\\
-7	0.0194014348089695\\
-6	0.0182536691427231\\
-5	0.0171233229339123\\
-4	0.0161610171198845\\
-3	0.0153657821938396\\
-2	0.0147437863051891\\
-1	0.0142878433689475\\
0	0.013903446495533\\
1	0.0135319698601961\\
2	0.0131789520382881\\
3	0.0129236057400703\\
4	0.0126932766288519\\
5	0.012509398162365\\
6	0.0122353350743651\\
7	0.011940548196435\\
8	0.0117380414158106\\
9	0.0115624405443668\\
10	0.011440347880125\\
11	0.0113559290766716\\
12	0.0113101359456778\\
13	0.0111987441778183\\
14	0.010988469235599\\
15	0.0108022894710302\\
16	0.0106191374361515\\
17	0.010459341108799\\
18	0.0103613054379821\\
19	0.0102615840733051\\
20	0.0101322159171104\\
};
\addlegendentry{Ours ($I_\textnormal{C}$=2)}

\addplot [color=mycolor9, dotted, line width=2.0pt, forget plot]
  table[row sep=crcr]{%
20	0.0101322159171104\\
30	0.00989924743771553\\
};
\addplot [color=mycolor9, only marks, mark size=2.0pt, mark=*, mark options={solid, fill=mycolor9, draw=mycolor9}, forget plot]
  table[row sep=crcr]{%
30	0.00989924743771553\\
};
\addplot [color=mycolor10, line width=2.0pt]
  table[row sep=crcr]{%
-20	0.0175589676946402\\
-19	0.0166011694818735\\
-18	0.0157231148332357\\
-17	0.014965838752687\\
-16	0.0142391137778759\\
-15	0.0135122993960977\\
-14	0.0129183102399111\\
-13	0.0124871218577027\\
-12	0.0120797231793404\\
-11	0.011678297072649\\
-10	0.011296521872282\\
-9	0.0109334699809551\\
-8	0.010615729726851\\
-7	0.0103760361671448\\
-6	0.0100635681301355\\
-5	0.00973053369671106\\
-4	0.0094646830111742\\
-3	0.00920397695153952\\
-2	0.00895331613719463\\
-1	0.00864145345985889\\
0	0.0083334855735302\\
1	0.00810854136943817\\
2	0.00792970973998308\\
3	0.00771782174706459\\
4	0.00749047845602036\\
5	0.00724907498806715\\
6	0.00703780446201563\\
7	0.00688794674351811\\
8	0.00671845022588968\\
9	0.00654818397015333\\
10	0.00634963251650333\\
11	0.00617388263344765\\
12	0.00599782867357135\\
13	0.00580662162974477\\
14	0.00561957946047187\\
15	0.00549668818712234\\
16	0.00534580461680889\\
17	0.00521346321329474\\
18	0.00506933266296983\\
19	0.00493845716118813\\
20	0.00481738662347198\\
};
\addlegendentry{Ours ($I_\textnormal{C}$=4)}

\addplot [color=mycolor10, only marks, mark=*, mark options={solid, fill=mycolor10, draw=mycolor10}, forget plot]
  table[row sep=crcr]{%
-20	0.0175589676946402\\
};
\node[fill=white, below, align=center, inner sep=0, font=\color{mycolor10}]
at (axis cs:-20,0.003) {0.018};
\addplot [color=mycolor10, only marks, mark=*, mark options={solid, fill=mycolor10, draw=mycolor10}, forget plot]
  table[row sep=crcr]{%
-10	0.011296521872282\\
};
\node[fill=white, below, align=center, inner sep=0, font=\color{mycolor10}]
at (axis cs:-10,-0.004) {0.011};
\addplot [color=mycolor10, only marks, mark=*, mark options={solid, fill=mycolor10, draw=mycolor10}, forget plot]
  table[row sep=crcr]{%
0	0.0083334855735302\\
};
\node[fill=white, below, align=center, inner sep=0, font=\color{mycolor10}]
at (axis cs:0,-0.007) {0.008};
\addplot [color=mycolor10, only marks, mark=*, mark options={solid, fill=mycolor10, draw=mycolor10}, forget plot]
  table[row sep=crcr]{%
10	0.00634963251650333\\
};
\node[fill=white, below, align=center, inner sep=0, font=\color{mycolor10}]
at (axis cs:10,-0.009) {0.006};
\addplot [color=mycolor10, only marks, mark=*, mark options={solid, fill=mycolor10, draw=mycolor10}, forget plot]
  table[row sep=crcr]{%
20	0.00481738662347198\\
};
\node[fill=white, below, align=center, inner sep=0, font=\color{mycolor10}]
at (axis cs:20,-0.01) {0.005};
\addplot [color=mycolor10, only marks, mark=*, mark options={solid, fill=mycolor10, draw=mycolor10}, forget plot]
  table[row sep=crcr]{%
30	0.00313499313779175\\
};
\node[fill=white, below, align=center, inner sep=0, font=\color{mycolor10}]
at (axis cs:30,-0.012) {0.003};
\addplot [color=mycolor10, dotted, line width=2.0pt, forget plot]
  table[row sep=crcr]{%
20	0.00481738662347198\\
30	0.00313499313779175\\
};
\addplot [color=mycolor10, only marks, mark size=2.0pt, mark=*, mark options={solid, fill=mycolor10, draw=mycolor10}, forget plot]
  table[row sep=crcr]{%
30	0.00313499313779175\\
};
\end{axis}
\end{tikzpicture}%
    \caption{Detection performance of coarse ToA estimation stage measured by the area under the receiver operating characteristic (ROC) curve for different signal-to-interference ratios (SIRs). These results are for an SNR of $30\,\text{dB}$.}
    \label{fig:auc_first_stage}
\end{figure}
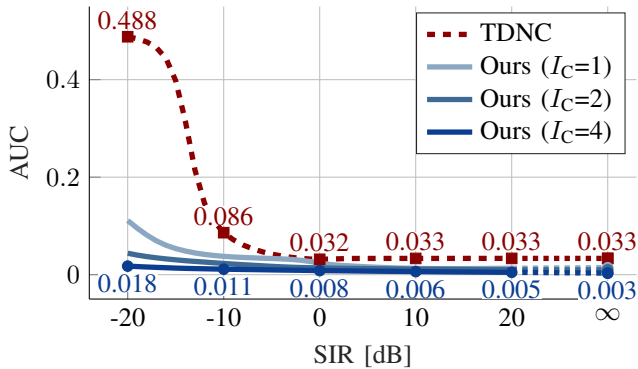

For the coarse ToA estimation stage, we evaluate the proposed algorithm for a fixed SNR of $30\,\text{dB}$, $I_\textnormal{C} \in \left\{1,2,4\right\}$. We compare our algorithm to the time-domain normalized correlation (TDNC) performed by most practical systems, which simply replaces the score function in~\fref{eq:coarse_toa_opt} by $\vecnorm{\bY_\ell\conj{\bmx}_{\textnormal{S},0}} / \frobnorm{\bY_\ell}^2$. The results are shown in Figure \ref{fig:auc_first_stage}.
Our solution outperforms TDNC at all SIRs and $I_\textnormal{C}$, showing a significant advantage for $\mathrm{SIR} \leq -10\,\mathrm{dB}$. Even at $\mathrm{SIR} = \infty$, where TDNC should theoretically achieve near-perfect performance, it still shows significant degradation, which reveals the impact of multipath propagation. Our algorithm mitigates this degradation, with detection performance improving as the rank approximation $I_\textnormal{C}$ increases.

\subsubsection{Fine ToA Estimation}
\label{sec:fine_toa_estimation_results}

\begin{figure}[tp]
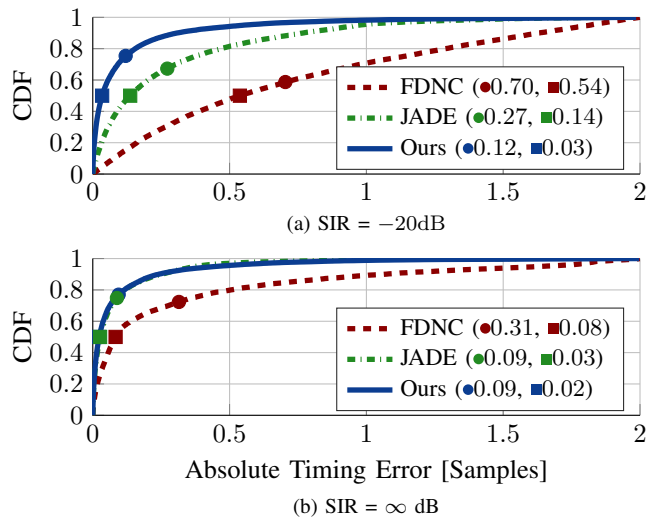

    \centering
    \begin{tikzpicture}
        \begin{groupplot}[
            group style={
                group size=1 by 2,
                vertical sep=32pt,
                xlabels at=edge bottom,
                ylabels at=edge left
            },
            width=\columnwidth,
            height=0.414\columnwidth
        ]
        \nextgroupplot[toacdf, ylabel={CDF}]
        \input{tikz/fine_sync_cdf_sir_-20_dB}

        \nextgroupplot[toacdf, ylabel={CDF}, xlabel={Absolute Timing Error [Samples]}]
        \input{tikz/fine_sync_cdf_sir_inf_dB}
        
        \end{groupplot}

        \node[anchor=north] at ([yshift=-12pt]group c1r1.south)
            {\footnotesize (a) SIR = $-20\mathrm{dB}$};

        \node[anchor=north] at ([yshift=-28pt]group c1r2.south)
            {\footnotesize (b) SIR = $\infty$ dB};
    \end{tikzpicture}
    \caption{Fine ToA estimation stage absolute timing error CDF for (a) a SIR of $-20\,\text{dB}$ and (b) a SIR of $\infty$. All results are shown for an SNR of $30\,\text{dB}$ and $I_\textnormal{F}=16$. The markers $\bullet$ denote the mean and $\blacksquare$ the median.}
    \label{fig:fine_toa_abs_err_cdf}
\end{figure}

\begin{figure}[tp]
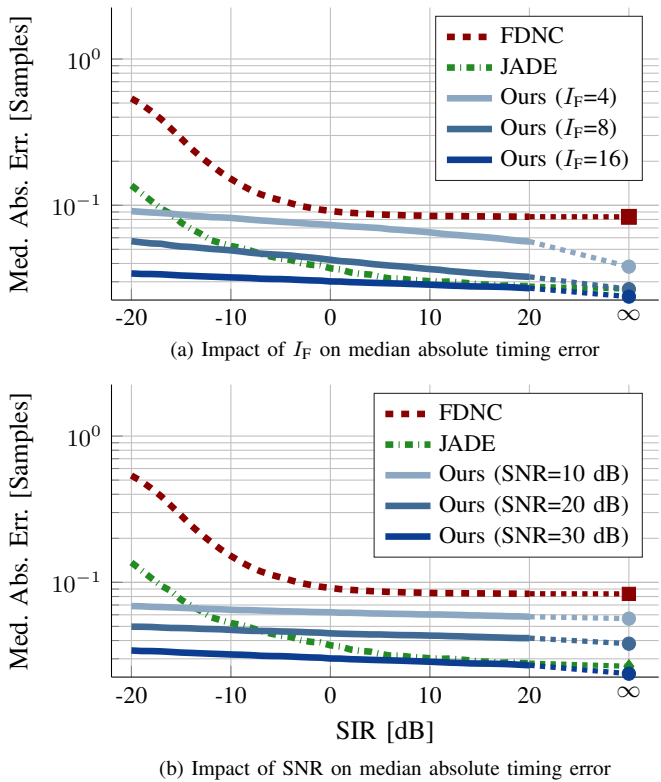

    \centering
    \begin{tikzpicture}
        \begin{groupplot}[
            group style={
                group size=1 by 2,
                vertical sep=32pt,
                xlabels at=edge bottom,
                ylabels at=edge left
            },
            width=\columnwidth,
            height=0.618\columnwidth
        ]

        \nextgroupplot[ablationtoa, ymode=log, ylabel={Med. Abs. Err. [Samples]}]
        \input{tikz/fine_sync_impact_of_I}
        \nextgroupplot[ablationtoa, ymode=log, xlabel={SIR [dB]}, ylabel={Med. Abs. Err. [Samples]}]
        \input{tikz/fine_sync_impact_of_snr}
        
        \end{groupplot}

        \node[anchor=north] at ([yshift=-12pt]group c1r1.south)
            {\footnotesize (a) Impact of $I_\textnormal{F}$ on median absolute timing error};

        \node[anchor=north] at ([yshift=-28pt]group c1r2.south)
            {\footnotesize (b) Impact of SNR on median absolute timing error};

    \end{tikzpicture}
    \caption{Ablation studies for fine ToA estimation. (a) shows the impact of $I_\textnormal{F}$, fixing the SNR to $30\,\text{dB}$, and (b) shows the impact of the SNR, fixing $I_\text{F}=16$. The curves for FDNC and JADE are also shown for reference for an SNR of $30\,\text{dB}$.}
    \label{fig:fine_toa_est_ablation_studies}
\end{figure}

For the fine ToA estimation stage, we evaluate the proposed algorithm for SNRs of $10\,\text{dB}$, $20\,\text{dB}$, and $30\,\text{dB}$, and $I_\textnormal{F} \in \left\{4,8,16\right\}$. As baselines, we consider the frequency-domain normalized correlation (FDNC), which replaces the score function in \fref{eq:fine_toa_opt} by $\big\|\tilde{\bY}_{\est{\ell}_{\textnormal{S},1}-1} \conj{\tilde{\bmx}}_{\textnormal{S}}\left(\tau'_1\right)\big\|^2/ \big\|\tilde{\bY}_{\est{\ell}_{\textnormal{S},1}-1}\big\|^2_\text{F}$, plus the subspace-based JADE algorithm from~\cite{VanDerVeen1997JADE}. 
FDNC represents a conventional approach for ToA estimate refinement, while JADE is a subspace method that extends the shift-invariance principles of ESPRIT to delay estimation. 
JADE targets high-resolution, computationally efficient estimation, making it a compelling baseline. For JADE, we evaluate all model-orders from $1$ to~$7$ and select the one with the minimum timing error. 
We also employ forward-backward averaging and spatial smoothing to handle correlated multipath sources.\footnote{We set $m_1=5$ and $m_2=4$ to maximize the number of identifiable MPCs. For the approximate joint matrix diagonalization step, we use the algorithm proposed in \cite{Cardoso1996JacobiAngles}, porting it to JAX. We only consider delays within the valid range between $0$ and $2T_s$ and take the lowest as the ToA estimate.}
Figure \ref{fig:fine_toa_abs_err_cdf} shows the CDF of the absolute timing error for the best-case scenario of $I_\textnormal{F}=16$, an SNR of $30\,\text{dB}$, and the two extreme cases for interference (SIR of $-20\,\text{dB}$ and no interference).
The proposed method shows a clear advantage over FDNC under both heavy interference and no interference, confirming the effective mitigation of multipath-induced bias. JADE outperforms FDNC but performs worse than our method in the low-SIR regime, where interference hampers channel estimation. 
We also analyze the impact of SNR and the chosen interference rank $I_\textnormal{F}$. For analyzing the first, $I_\text{F}$ is set to $16$; for the converse case, the SNR is fixed to $30\,\text{dB}$. The results are shown in Figure \ref{fig:fine_toa_est_ablation_studies}.
The estimation error decreases with increasing SNR and $I_{\textnormal{F}}$. Higher SNR enables our algorithm to better estimate spatial signatures, while higher $I_{\textnormal{F}}$ enables it to suppress more nuisance MPCs. Importantly, performance improves as $I_{\textnormal{F}}$ increases, indicating that a precise rank estimate is not required. This stands in contrast to JADE, which incurs in severe ToA estimation errors if the number of MPCs is misspecified.
\begin{table}[tp]
\centering
	\renewcommand{\arraystretch}{1.2}
	\caption{Runtimes of fine ToA estimation.}
    \label{tbl:fine_toa_runtimes}
	\small
	\begin{tabular}{@{}ll*{4}{c}@{}}
		\toprule
		\multirow{3}{*}{Method}  & \multirow{3}{*}{SNR}  & \multicolumn{4}{c}{Runtime [ms]} \\
		\cmidrule(lr){3-6} 
		& & Mean & Max. & Min. & Std. \\
		\midrule
		\multirow{3}{*}{Ours} & $30\,\text{dB}$ & $8.19$ & $11.39$ & $5.43$ & $1.55$ \\
        & $20\,\text{dB}$ & $7.81$ & $10.61$ & $3.69$ & $1.36$ \\
        & $10\,\text{dB}$ & $7.78$ & $10.58$ & $2.87$ & $1.42$ \\
        \midrule
	    \multirow{3}{*}{FDNC} & $30\,\text{dB}$ & $0.56$ & $0.81$ & $0.38$ & $0.10$ \\ 
        & $20\,\text{dB}$ & $0.53$ & $0.75$ & $0.37$ & $0.08$ \\ 
        & $10\,\text{dB}$ & $0.52$ & $0.73$ & $0.37$ & $0.07$ \\ 
        \midrule
        \multirow{1}{*}{JADE} & $10\,\text{dB}\textnormal{-}30\,\text{dB}$ & $17.02$ & $69.16$ & $2.98$ & $18.45$ \\
 
	\bottomrule
	\end{tabular}
\end{table}

\subsubsection{Runtime Comparison}
Finally, \fref{tbl:fine_toa_runtimes} compares algorithm runtimes for an AMD Ryzen 9 7900X CPU and $128$\,GB of RAM. As expected, FDNC is the fastest: although the number of score function and corresponding derivative evaluations is the same on average as for our method, they are much cheaper in FDNC's case, which does not involve an EVD. Even so, our algorithm's maximum runtimes stay below JADE's\footnote{Measured only for the model-order yielding the minimum error.}, and are still appropriate for many practical applications, including wireless time transfer and positioning, which usually require update rates up to $100$\,Hz.\footnote{In our implementation, we compute the full signal covariance EVD for building the spatial filter. Runtimes could be further reduced by employing orthogonal iterations to only update the required $I_\textnormal{F}$ singular vectors between score function evaluations. This would also allow $I_\textnormal{F}$ to be tuned for the desired error/runtime trade-off. Such improvements are, however, left for future work.}
\section{Conclusions}
\label{sec:conclusions}

We have proposed a novel ToA estimation method that utilizes multiple receive antennas to resolve multiple propagation paths beyond the Rayleigh limit under strong interference. Unlike prior work, our method jointly addresses interference, multipath propagation, and high-resolution estimation without auxiliary model-order estimation and with tractable computational complexity.
Our results demonstrate superior performance to correlation-based methods and surpass subspace-based methods in low-SIR regimes, improving upon the latter's latency and model-order sensitivity.

\balance

\bibliographystyle{IEEEtran}
\bibliography{bib/VIPabbrv,bib/confs-jrnls,bib/publishers,bib/VIP_190331,bib/bibliography}

\end{document}